**Splay-induced charging of ferroelectric droplets in highly non-uniform electric fields**

Lorenzo Fiorentini[1], Elia Rocchetti[1], Raouf Barboza[1], Roberta Galeazzi[2], Liana Lucchetti[1*]

1. *Dipartimento SIMAU, Università Politecnica delle Marche, via Brecce Bianche, 60131 Ancona, Italy. E-mail: l.lucchetti@univpm.it*
2. *Dipartimento DISVA, Università Politecnica delle Marche, via Brecce Bianche, 60131 Ancona, Italy.*

**Abstract**

Ferroelectric sessile droplets on a superhydrophobic surface can be put into motion by highly non-uniform electric fields, in a direction that depends on the field sign. This indicates that droplets carry a non-vanishing electrostatic charge, which we interpret as acquired through contact electrification at the interface with the superhydrophobic substrate. This phenomenon is not observed in the presence of a uniform or quasi-uniform electric field, suggesting that droplets charging is favored in highly nonuniform fields able to induce localized distortions of the ferroelectric polarization, in turn producing localized charge accumulation sites. Electrostatic charging is thus understood as a splay-induced contact electrification and interpreted as a strategy to minimize the accumulated charge. Electrophoretic-like motion is observed in about 30% of the experiments, meaning that splay-induced charging requires the proper combination of conditions such as droplets size, distance from the field source, freshness of the substrate coating and other experimental details that are not easily quantified. When charging does not occur, effects related to droplet polarization such as dielectrophoresis, ejection of polarized fluid jets and electrostriction-like deformations are observed.

**Introduction**

The recently discovered ferroelectric nematic ($N_F$) phase[1-4] represents the first ferroelectric fluid with a workable viscosity and its peculiar combination of fluidity and polarity has opened the gate to many new phenomena. This phase is the polar version of the well-known nematic liquid crystal phase, where the direction of the mean molecular orientation is expressed by the nematic director **n**. Even when formed by molecules carrying a dipole moment, the nematic phase is nonpolar, i.e. molecular dipoles are oriented with equal probability in the two opposite directions, thus making the states **n** and - **n** indistinguishable. On the contrary, the molecular arrangement in $N_F$ breaks the head-tail symmetry of the director, making the phase macroscopically polar with the polarization vector **P** approximately collinear with **n**. This feature leads to many peculiar properties, among which those related to the polar coupling of these materials to electric fields, very different from the dielectric coupling of the conventional nematic phase[5-10]. One of the most intriguing properties of the $N_F$ phase, which is a consequence of the combination of large spontaneous polarization and fluidity, is the readiness with which it can accumulate polarization charges at any interface with a dielectric medium, either solid walls or air, by small collective rotations of **n** and thus of **P,** to cancel any electric-field component normal to the dielectric surfaces[11]. Indeed, any splay deformation of the molecular ordering leads to accumulation of bulk polarization charges of density $\rho = -\nabla \cdot \mathbf{P}$, in turn producing internal electric fields. Such effective screening, which has been called "fluid superscreening" is achieved through minor local changes in the orientation of **P** and is the leading and fastest component of the multiscale dynamic response of $N_F$ to electric fields, determining all the subsequent time evolution[11].

When the electric field acting on $N_F$ is generated either by the pyroelectric or photovoltaic effect of a ferroelectric solid crystal, additional peculiar behaviors have been observed spanning from Rayleigh-like electromechanical instability[12] to optical control of the emitted fluid jets[13] and of the polar topology[14] and to light-induced walking of ferroelectric droplets[15].

In this work we keep on studying the unique features of the polar coupling to electric fields of the $N_F$ phase in the case where the field is generated by ferroelectric solid surfaces. Specifically, we analyze the behavior of sessile droplets in proximity to pyroelectrically charged lithium niobate (LN) crystals in conditions of minimized surface pinning, with the main aim of controlling their motion. Droplets with diameters of the order of 100 μm are deposited on a Polytetrafluoroethylene (PTFE) coated glass substrate, positioned either close to a single LN crystal or between two LN crystals arranged with parallel or antiparallel polar axes (see Fig. 1 and details in Materials and Methods). The fringing field generated by the LN ferroelectric crystals is non-uniform and acts on ferroelectric droplets causing either their deformation, possibly followed by electromechanical instability, or their motion. The force responsible for droplets' motion is attractive or repulsive, depending on the sign of the fringing field, indicating that the dominant interaction is of electrophoretic type, in turn suggesting that droplets become electrically charged after the transition to the $N_F$ phase. Interestingly, the same droplets do not move under a uniform or quasi-uniform electric field, meaning that droplet charging occurs only in the presence of a highly non-uniform electric field. We understand this phenomenon as a splay-induced release of charges that gives rise to contact electrification at the interface with the PTFE-coated substrate. Charge release appears as a strategy that droplets use to minimize the accumulated electrostatic energy, alternative to the emission of polar fluid jets observed when they lie directly on LN bare crystals[12].

Our findings shed light on the behavior of highly polar fluids in non-uniform electric fields. As we shall show in this manuscript, $N_F$ droplets can behave as charged objects or as neutral, polar fluids depending on the field profile and on the specific combination of experimental parameters such as droplet size, initial distance from the field source and quality of the PTFE coating.

The possibility of controlling the motion of ferroelectric liquid droplets in proximity to ferroelectric crystals might have high impact in the field of microfluidics, where the manipulation of complex fluids is still widely unexplored. The motion of droplets of isotropic fluids induced by the photovoltaic and pyroelectric charging of ferroelectric crystals have been recently described and the system has been proved to be a good candidate for the realization of microfluidic platforms[16-18]. Compared to these works, we consider here anisotropic, polar fluids, and use a different experimental geometry where droplets are not in direct contact with the ferroelectric solid. On the one hand, this gives us additional degrees of freedom in controlling the spatial profile of the fringing field, which largely affects the droplets' behavior; on the other, our system can easily be integrated into existing glass or PDMS microfluidic platforms, without the need to build new chips directly on LN crystals.

**Materials and methods**

The ferroelectric liquid crystal used in this work is 4-[(4-nitrophenoxy)carbonyl]phenyl2,4-dimethoxybenzoate (RM734). It was synthesized as described in ref. [1] and its structure and phase diagram have already been reported[1,19]. In this compound the ferroelectric nematic phase appears through a second order phase transition when cooling from the conventional higher temperature nematic (N) phase and exists in the range 133°C < T < 80°C[1,19]. The spontaneous polarization **P** of RM734 is either parallel or antiparallel to the molecular director **n,** defining the average orientation of the molecular axis, and exceeds 6 μC/cm$^2$ at the lowest T in the $N_F$ phase[1].

The lithium niobate (LN) ferroelectric substrates are 900 μm thick undoped z-cut crystals (PI-Kem), having the polar axis orthogonal to the slab surfaces. The bulk spontaneous polarization $P_{LN}$ of LN crystals along the [0001] z-axis is 70 μC/cm$^2$ and does not depend significantly on temperature in the explored range since its Curie temperature is much higher (≈ 1140°C). Because of very efficient compensation mechanisms at the z-cut surfaces, the equilibrium surface charge is instead only about $10^{-2}$ μC/cm$^2$ [20]. This value significantly increases when temperature variations are induced thanks to the pyroelectric effect[21-23], a transient phenomenon observable during and shortly after the variation[21] and due to the slow free charge relaxation in LN. The pyroelectric coefficient of undoped LN crystals is of the order of $10^{-4}$ C/m$^2$K at room temperature[24] and increases by one order of magnitude around 100°C[25]. Given the temperature used in our experiments, dictated by the RM734 phase diagram, we can thus expect an induced surface charge density of the order of 1 μC/cm$^2$, for temperature variations of a few degrees ramped in a short time compared to the LN charge relaxation.
The sign of this surface charge depends on the direction of the polar axis and on the sign of temperature variation. Specifically, at the top surface of z-cut LN crystals with positive axis (i.e. directed from the bottom to the top surface), the uncompensated charge is positive on cooling and negative on heating; the contrary happens reverting the crystal, i.e. when the polar axis is negative[23,26,27]. Due to the finite size of LN crystals, the pyroelectric charging produces a fringing electric field $E_f$ external to the crystal itself, which is a fraction *f* of the internal field $\sigma_{LN}/\varepsilon_0$, with $f \approx 10^{-3}$ depending on the crystal size and geometry[28].The spatial distribution of $E_f$ in our experimental conditions, is reported in Fig. 1.

The RM734 droplets used in our experiments have a diameter ranging between 70 μm and 300 μm as measured with a calibration slide and were obtained as described in the SI (Note 1). They were deposited on 100 μm thick glass coverslips treated by a Polytetrafluoroethylene (PTFE) superhydrophobic coating (SI Note 2), which guarantees minimum pinning of the sessile RM734 droplets[10]. Before droplets deposition, the coverslip was heated to T = 170°C, corresponding to the RM734 nematic phase. Then, it was positioned on top of two equally thick substrates arranged side by side and separated by 1 mm gap, as shown in Fig. 1a-c'. The two substrates are either one LN crystal and one glass slide (Fig. 1 a), or two LN crystals arranged with parallel or antiparallel polar axes (Fig. 1b and c). The fringing field in the region where droplets are positioned is shown in Fig. 1, panels a''-c''', for the three configurations considered in this work.
RM734 droplets are positioned on the coverslip at various distances with respect to the border of LN crystals (marked by a red dashed line in Fig. 1 a'-c') along an imaginary line that cuts the crystals into two equal parts (green dashed line). In these positions, the y component of $E_f$ can be neglected due to symmetry considerations. Successively, temperature was slowly lowered to 105°C, which is in the ferroelectric range of RM734, with an average cooling rate of 0.1°C/s. At lower temperatures, the increase of viscosity close to the RM734 crystalline phase hinders both droplets motion and deformation. Heating cycles were also considered. After being cooled down to 105°C, the system was reheated to 165°C. Droplets keep their $N_F$ phase up to the transition to the N phase, that we observe at T = 150 °C. Due to the characteristics of our temperature controller, the heating rate is on average one order of magnitude faster than the cooling rate.
The behavior of $N_F$ droplets has been monitored with the aid of a polarizing optical microscope (POM) and a CCD camera operating at 25 fps.

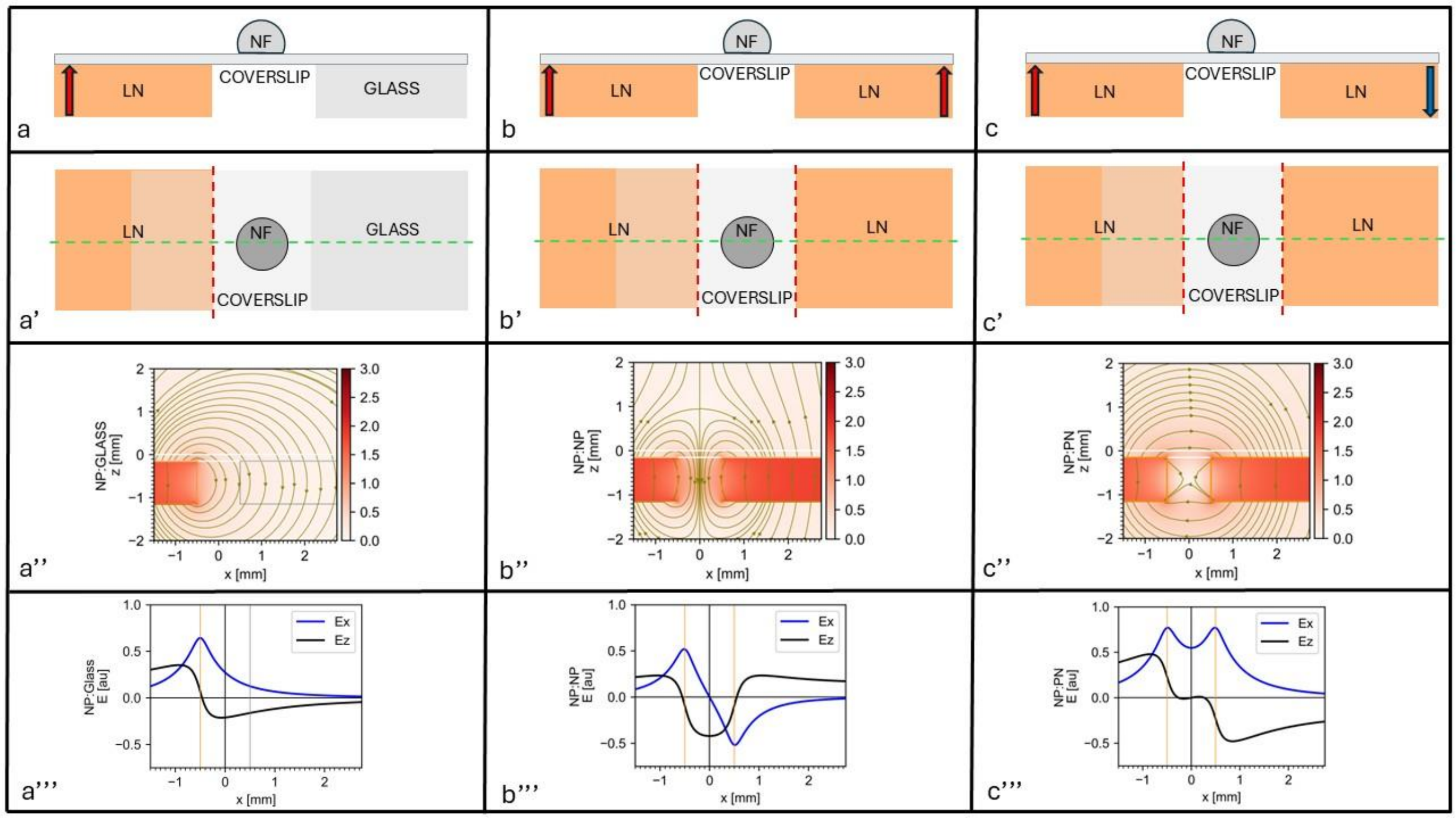

**Figure 1:** Experimental geometry and profile of the fringing field. Panels a-c: experimental arrangement, sketch of the side view. a) single LN crystal; b) two LN crystals with parallel polar axes (positive in the figure); c) two LN crystals with antiparallel polar axes. The LN polar axis is indicated by red (positive) and blue (negative) vertical arrows. The $N_F$ droplets lie on the thin coverslip coated by PTFE and are positioned in correspondence of the gap between the LN crystal and a glass substrate (a) or between the two LN crystals (b and c). Panels a'-c': top view of the same arrangements. The dashed green line represents the middle of the LN crystals. Along this line, the y component of the fringing field is zero by symmetry. Red dashed lines mark the border of LN crystals. Panels a''-c'') Field lines, side view. Colored slabs represent LN crystals, the thin white slab on top of them represents the PTFE-coated coverslip. The fringing field is strongly non uniform in the region that will be occupied by $N_F$ droplets, except for the case in panel c'') where the vertical component of the field is extremely small almost everywhere and the horizontal component is quasi uniform. Panels a'''-c''') x and z components of $E_f$ for the three configurations, as a function of the position along the x axis. Yellow vertical lines mark the borders of LN crystals, while the grey vertical line in panel a''' represents the border of the glass substrate positioned below the coverslip to maintain it horizontal.

## Results and discussion

A rough estimation of the low wettability of the PTFE coated glass surface is given by the contact angle of RM734 sessile droplets, which exhibits an average value of 97° in the N phase and 89° in the $N_F$ phase (from measurements on ten droplets in tilted configuration. SI Note 3). These values can be compared to those obtained for the static contact angle, in the case of RM734 droplets on fluorolink coated substrates: 70° in the N phase and 55° in the $N_F$ phase[15]. Sessile $N_F$ droplets on fluorolink-coated LN substrates can be put into motion with an average speed of a few μm/s, by irradiating the crystal with a gaussian beam having intensity of the order of $10^2$ W/cm$^2$ [15].

In the experiments reported here, droplets are instead not in direct contact with LN crystals and their motion results from the interaction with the fringing field generated by pyroelectric charging. Droplet motion is observed in approximately 30% of the attempts and only in conditions of fresh PTFE coating. Droplets deformation and electromechanical instability events cover the remaining 70%.

When motion occurs, the typical behavior of $N_F$ droplets when they are in proximity to a single LN crystal (as in Fig. 1a), is summarized in Fig. 2, where four different experimental situations are represented, obtained using LN with positive or negative polar axis subjected to cooling or heating cycles. Motion is observed only in the $N_F$ phase and when droplets move, they roll across the surface along the green horizontal line drawn in Fig. 1 a'. Droplets typically stop at the end of the cooling cycle, and their motion is reestablished upon heating the system. Motion is faster upon heating, due to the faster rate of temperature variation which results in a higher fringing field. Figure 2 has been extracted from Videos S1 and S2. Both videos have been sped up with respect to the original ones. For this reason, the times reported in Fig. 2 do not correspond to those in the Videos.

Surprisingly, the direction of motion depends on both the sign of the LN polar axis (indicated as + or – in the figure) and of the temperature variation. Specifically, droplets move toward the LN crystal when the polar axis is negative and the system is cooled down (Fig. 2a), or when the polar axis is positive and the system is heated up (Fig. 2a'''). However, when the sign of the temperature variation is reverted, droplets move in the opposite direction, that is they walk away from the crystal in case of negative polar axis upon heating (Fig. 2a') or in case of positive polar axis upon cooling (Fig. 2a''). Interestingly, droplets move faster when LN crystals with negative polar axis are used.
Keeping in mind that the top surface of LN with positive polar axis exhibits positive charge on cooling and negative charge on heating, while the contrary happens reverting the crystal, our observations can be summarized stating that $N_F$ droplets are always repelled by positive charges and always attracted by negative charges, i.e. they behave as if they were positively charged.

During the cooling phase, droplets typically stop a few tenths of seconds after the end of the cycle due to the relaxation of the fringing field $E_f$ (see Video S2). The field relaxation depends on the restoration of screening charges through gas molecules ionization due to the exposure to air and is characterized by extremely short response time[26]. We understand the relatively slow droplet reaction to the fast relaxation of $E_f$, as due to a small contribution of droplet inertia. In some experiments however, the motion ends during the cooling cycle, apparently because droplets meet regions of large pinning due to non-uniform PTFE coating. The presence of areas characterized by large pinning is suggested by two observations: i) droplets that stop during the cooling cycle often undergo a small deformation in the direction of motion and eventually emit tiny jets; ii) droplets appear sometimes slowed down over short periods of time, as if they were sticked on the surface (see Video S1).
During the heating phase, droplets stop either due to electromechanical instability events, which are more common and violent on heating than on cooling (see for example Videos S3-1 and S3-2), or because of the transition to the non-polar nematic phase (as in Video S2). Indeed, after the transition droplets slow down and eventually stop, even if the heating cycle is not finished, suggesting that they possibly lose their charge.

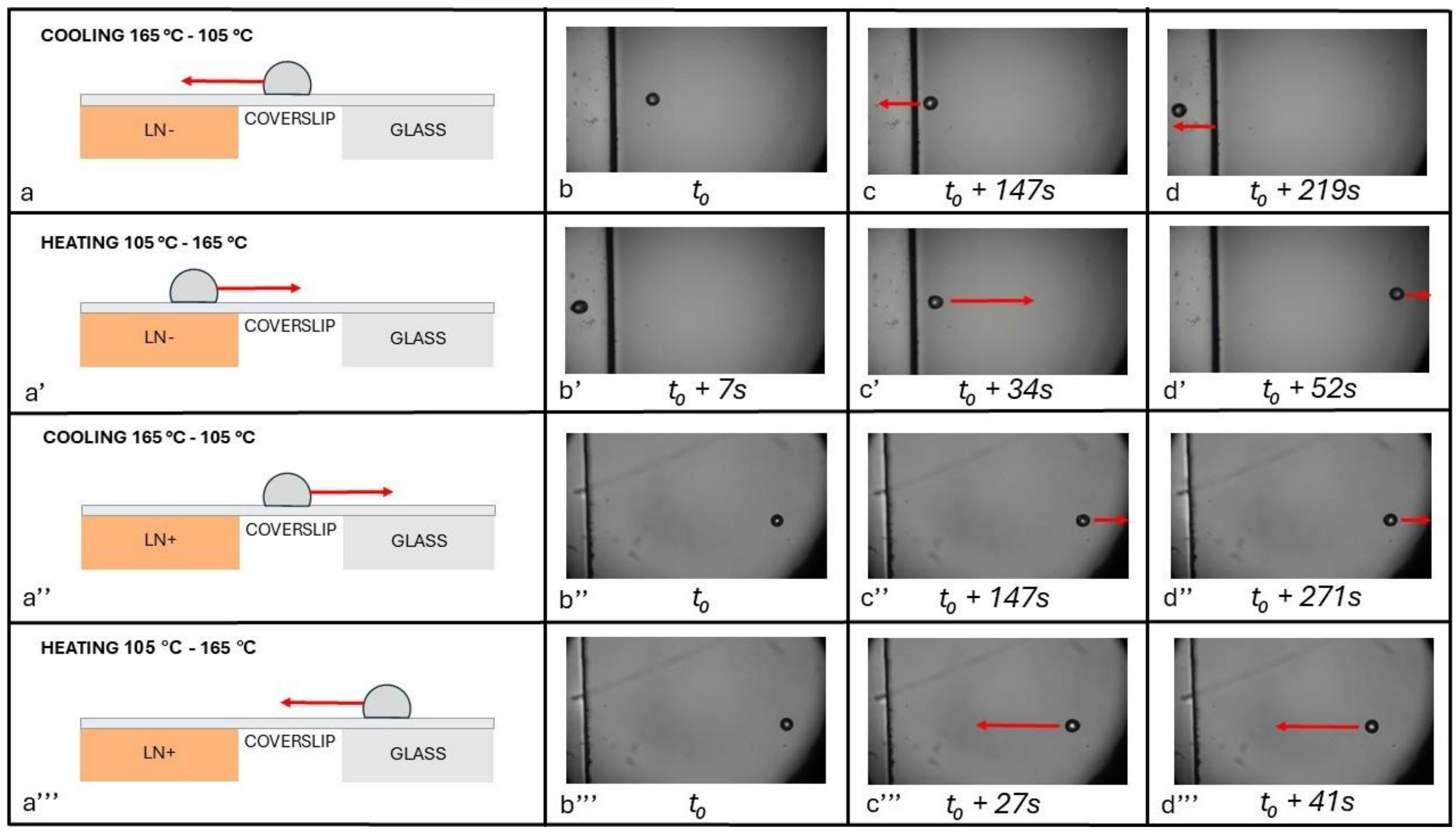

**Figure 2:** Typical behavior of moving droplets exposed to the fringing field of a single LN crystal. a - a''') Sketches of the experimental arrangement in case of droplets attraction (a and a''') and repulsion (a' and a''). LN- and LN+ indicate LN with negative and positive polar axes, respectively. b,c,d) Video frames taken at different instants showing an RM734 $N_F$ droplet moving toward the LN- crystal upon cooling. b',c',d') Video frames taken at different instants showing an RM734 $N_F$ droplet moving away from the LN- crystal upon heating. b'',c'',d'') Video frames taken at different instants showing an RM734 $N_F$ droplet moving away from the LN+ crystal upon cooling. b''',c''',d''') Video frames taken at different instants showing an RM734 $N_F$ droplet moving toward the LN+ crystal upon heating. $t_0$ marks the beginning of cooling or heating cycle. Panel b' shows the droplet 7 s after $t_0$, since at $t_0$ it is still out of the field of view. Motion is faster upon heating. Figure 2 has been extracted from the original videos. Videos S1 (LN-) and S2 (LN+) in the SI are the accelerated versions of the original ones.

The hypothesis of charged droplets is further supported by other experimental evidence. Additional experiments performed by placing RM734 droplets above or below the imaginary green line in Fig. 1a', where the y component of $E_f$ is no longer negligible, show that they move along a diagonal direction, following the field lines (see Video S4). Moreover, in some of our measurements, $E_f$ undergoes several fluctuations and possible changes of sign, which are observable as birefringence fluctuation of the LN crystal, likely connected to instability of the temperature controller and/or to piezoelectric induced LN deformations. These fast field variations affect the behavior of $N_F$ droplets by hindering the motion and inducing back and forth deformation of the rim (see Video S5).

However, when we tested the behavior of $N_F$ droplets under a uniform electric field (details in the SI), we did not observe any motion. In our experimental conditions, the fringing field generated by pyroelectric charging is $10^{-3}$ times the internal pyroelectric field[12], i.e. not higher than $10^5$ V/m [29]. We applied a static field E = 5 x $10^5$ V/m and no motion was observed. Then, we did an additional test using an LN crystal as a charge generator (see SI Note 4) and again $N_F$ droplets did not move. The only effect of the uniform field was to produce droplets deformation, due to the polarization of the polar fluid under the applied electric field.

Despite this experimental evidence, droplets charging appears the most reasonable explanation for the observed motion. An apo-electric response characterized by a dielectrophoretic force directed oppositely to the region of strongest field[30], could in principle be considered but it is easily ruled out by simple observations (see SI Note 5, 6 and Video S6). The experimental scenario is thus that of droplets that move only when they are in the $N_F$ phase and only when they are exposed to a non-uniform electric field. Assuming electrophoretic motion, this means that droplets acquire a net electric charge just in the $N_F$ phase and just under the action of a non-uniform field.

One possible charging mechanism of sessile droplets is contact electrification[31], which has been recently reported for water droplets on superhydrophobic surfaces such as Fluorinated polydimethylsiloxan[32] and PTFE[33]. In this latter study, when a water droplet hits the fresh PTFE layer, electron transfer takes place from water to PTFE resulting in the droplet becoming positively charged. Electron transfer, identified as the dominant mechanism of contact electrification[31], is attributed to the overlap of electron clouds between two atoms or two molecules under the influence of liquid pressure[33].
In the case of $N_F$ droplets, we propose the following scenario. In the presence of the fringing field the ferroelectric droplet becomes polarized. This takes place through a small reorientation of **P** by an angle such to deposit polarization charge on the droplet surfaces that cancel the internal field. This process produces splayed regions where bulk polarization charges accumulate[11,12]. When droplets are in direct contact with LN crystals, the corresponding electrostatic energy is minimized by expelling polarized fluid jets with charged tips, as reported in ref. [12] and sketched in Fig. 3a[12]. We understand the droplet charging observed here as an alternative strategy to relax the accumulated electrostatic energy, favored by the contact with the superhydrophobic PTFE layer, which easily acts as an electron acceptor[33] (a considerably high negative charge affinity of -190 nC/J has been reported for this fluorinate polymer[34]). When $N_F$ droplets accumulate bound charges in defected areas, electrons can be released to diminish the corresponding accumulated energy leaving the droplet positively charged, as shown in Fig. 3b.
Electron transfer is favored for temperature values largely exceeding room temperature[31], as in our experiments. In the case of RM734 droplets into contact with the PTFE, the increased electron cloud overlap might result from a combination between the decrease of the contact angle observed upon entering the $N_F$ phase and the tiny deformation localized in the splayed area and due to the competition of electrostatic energy and surface tension. With this in mind, we expect the process to require the formation of highly localized splayed regions - resulting in high local density of accumulated charge - favored by our experimental arrangement, where the LN fringing field is strongly non uniform on the scale of the average droplet dimension. In this scenario, droplets do not move under a uniform field since they simply do not charge. Indeed, a uniform electric field would give rise to a more uniform droplet polarization, resulting in distributed splay. This is confirmed by recent observations on $N_F$ droplets confined between two LN substrates to form a sandwich cell, aimed at assessing the correlation between topological defects and electromechanical instability[35]. In this configuration droplets are positioned at the center of the LN crystals, where the field is almost uniformly vertical on the microscale. The general tendency of the system is to keep the droplet rim as electrically uniform as possible, avoiding the formation of local charge accumulation sites[35]. As an additional remark, we highlight that the instability events observed in the present work are much rarer compared to those observed in ref. [12] and always preceded by droplet deformations localized in front of the LN border, indicating localized charge accumulation sites (see Videos S3-1, S3-2, and Fig. S9).
Contact electrification as a spontaneous process that lowers the electrostatic energy of the whole system, has recently been proposed also for polymers in contact with PTFE, and the electrostatic energy release has been used to quantify the tendency for charge transfer[36]. The larger the electrostatic energy released, the higher the tendency for charge transfer. This quantity is reminiscent of the work function difference - the difference between the energy required to extract an electron from an atom or a molecule - which governs the amount and direction of charge transfer between two contacting materials[36]. The work function of RM734 has been calculated with the aid of the Density Functional Theory Analysis (SI, Note 7). Its value of 8.47 eV, is lower than the 11.0 eV of PTFE[36], indicating that electrons can be transferred from RM734 molecules to PTFE, in agreement with the scenario we propose.
The electrostatic energy that a RM734 droplet on a LN substrate has to accumulate to overcome the surface tension and undergo an instability event can be estimated based on ref. [12] to be $U = \frac{k_C q Q}{R}$, where $k_C$ is the Coulomb constant, $R$ is the droplet radius, $Q = \sigma A$ is the droplet charge and $A$ its surface and $q = 2PS$ is the charge accumulated in a defect of area $S$. Assuming $S$ = 0.2 $\mu m^2$ [12] and $R$ = 100 μm one obtains $U$ of the order of 1 pJ. Considering the droplet as a hemisphere and using the RM734 mass density $\rho_m$ = 1.3 g/cm$^3$ [37] and molecular weight M = 423 g/mol[38], we obtain the number of moles $n$ corresponding to the mass of one droplet with $R$ = 100 μm as $n = 6 \times 10^{-9}$, meaning that U corresponds to about 0.2 kJ/mol. This value is nicely comparable to the electrostatic energy release of the order of 0.1 kJ/mol calculated in ref. [36] for materials with work functions comparable to those at play here.
The maximum number N of RM734 molecules involved in the electrification process can be roughly estimated considering that if every released electron costs 8.47 eV, that is about 13.5 x $10^{-19}$ J, 1 pJ corresponds approximately to N = $10^6$, which is a very tiny fraction of the total number of molecules $N_{TOT}$ derived from the number of moles: $N_{TOT}$ = 3.6 x $10^{15}$. This fraction is reasonably related to RM734 molecules closer to the interface with PTFE and to the defect area. Note that this value of N corresponds to the scenario in which all the energy stored by the droplet is released in this way, which is not necessarily the case: some energy might remain and result in droplet fission when it becomes higher than surface tension.

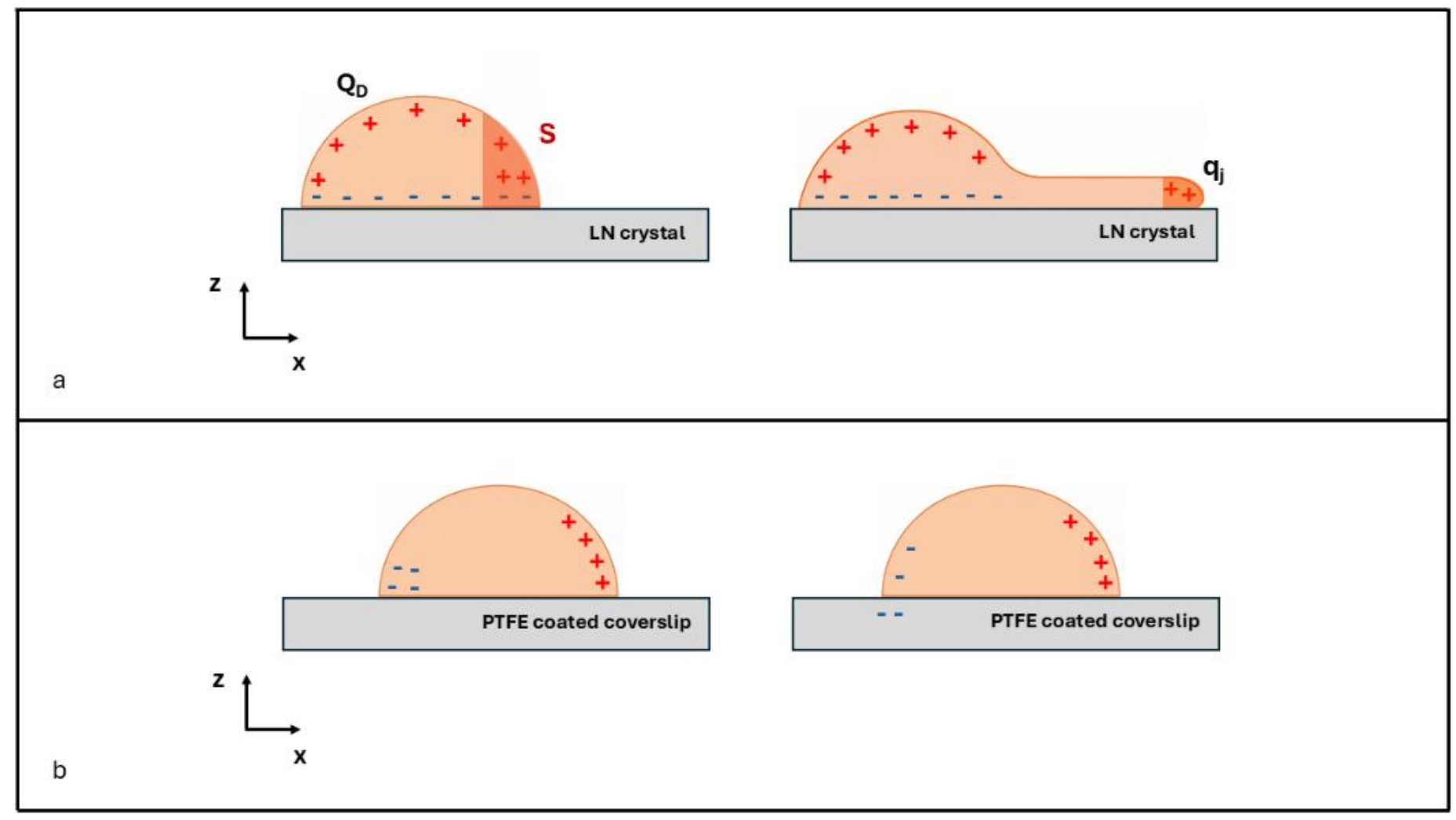

**Figure 3:** Comparison between ejection of polar fluid jets (a) and release of electrons proposed here as an alternative strategy to minimize electrostatic energy (b). a) Possible configuration leading to accumulation of charges at the droplet rim, in the case of $N_F$ droplets lying on the center of pyroelectrically charged uncoated LN substrates, where the field is largely vertical. b) Possible arrangement of the droplet polarization that could lead to localized regions of charge accumulation and consequent release of electrons to the PTFE-coated glass substrate, in the case of $N_F$ droplets positioned close to an LN charged crystal.

A closer look at droplets' motion can give us additional information on the proposed splay-induced charging.

Figure 4 shows examples of the curves describing droplet motion in the same experimental conditions reported in Fig. 2, for droplets of different size and different initial distance from the LN border. In these graphs the x axis corresponds to the green dashed line in Fig. 1a' and motion occurs only in this direction. The time t = 0 represents the beginning of cooling or heating processes. The value x = 0 marks the border of LN crystals; x values can be either positive or negative depending on the position of LN with respect to RM734 droplets (left or right). Distances are measured from the droplet center.

Although droplets all behave in the same way – all of them are attracted by LN crystals with negative (positive) polar axis on cooling (heating) and repelled reverting the sign of temperature variation – the curves describing droplet motion exhibit very different slopes and onset times. This indicates that the interaction with the fringing field, which always occurs in the $N_F$ phase, has different delays with respect to the N/$N_F$ transition and different strength depending on the specific measurement. As a general trend, the steepest curves correspond to droplets large compared to the scale of spatial non-uniformity of the field, which is in line with our interpretation: the more non-uniform is the field, the more localized are the induced splay deformations, which in turn promote contact electrification through electrons release. The different shapes of the curves can thus be ascribed to different amounts of droplet charging and to distinct times required to develop distorted areas. The behavior of each droplet is affected by size and initial distance from LN, and possibly by other experimental parameters that are not easily quantified, such as fluctuations of the temperature controller, quality of PTFE coating, convective motion of the fluid during temperature variation, pyroelectricity and piezoelectricity of the polar fluid.

Some of the curves pass through x = 0, confirming that droplets can travel above the LN crystal for quite long distances, and that the LN border, i.e. the region of maximum field gradient, does not represent the equilibrium position, as one would expect in case of neutral objects. Droplets are indeed attracted or repelled by the charged crystal surface, rather than by the region of maximum field.

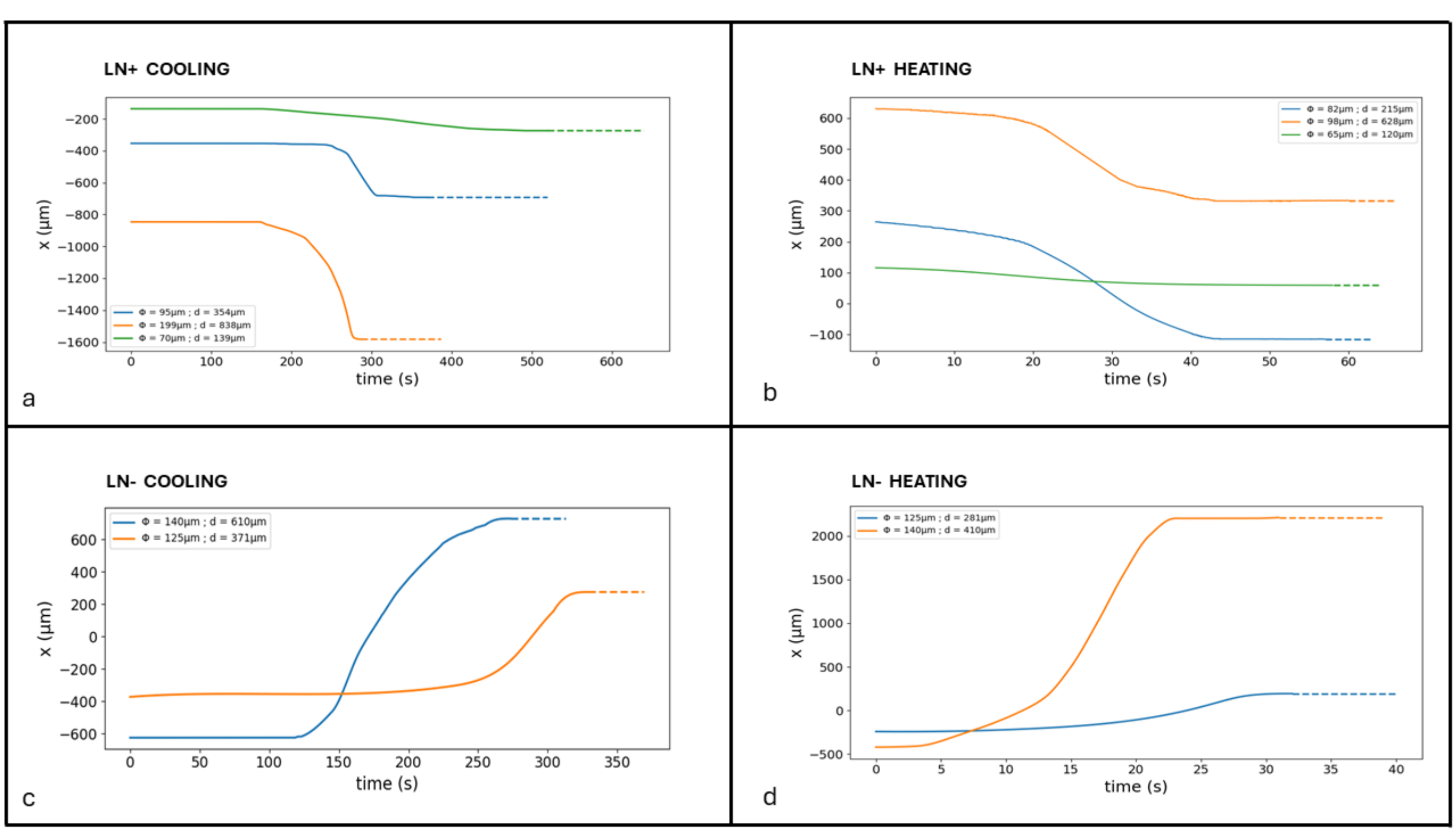

**Figure 4:** Droplet distance from the border of LN crystals (x = 0) as a function of time, for positive (a,b) and negative (c,d) polar axis, during cooling and heating cycles. LN- and LN+ indicate the sign of the polar axis. The time t = 0 s corresponds to the beginning of cooling or heating processes. Different colors indicate different droplets, with different size and different initial distance from LN crystals, as reported in the legend. The dashed parts of the curves represent regions were droplets exit the field of view.

Droplets' velocity, obtained as the time derivative of the curves in Fig.4, is shown in Fig. 5. If, to a very rough approximation, we neglect droplets inertia and surface pinning, the force exerted by $E_f$ is balanced by the viscous friction due to internal fluid motion, which we assume proportional to droplets velocity in agreement with experiments on $N_F$ droplets moved by light[15].

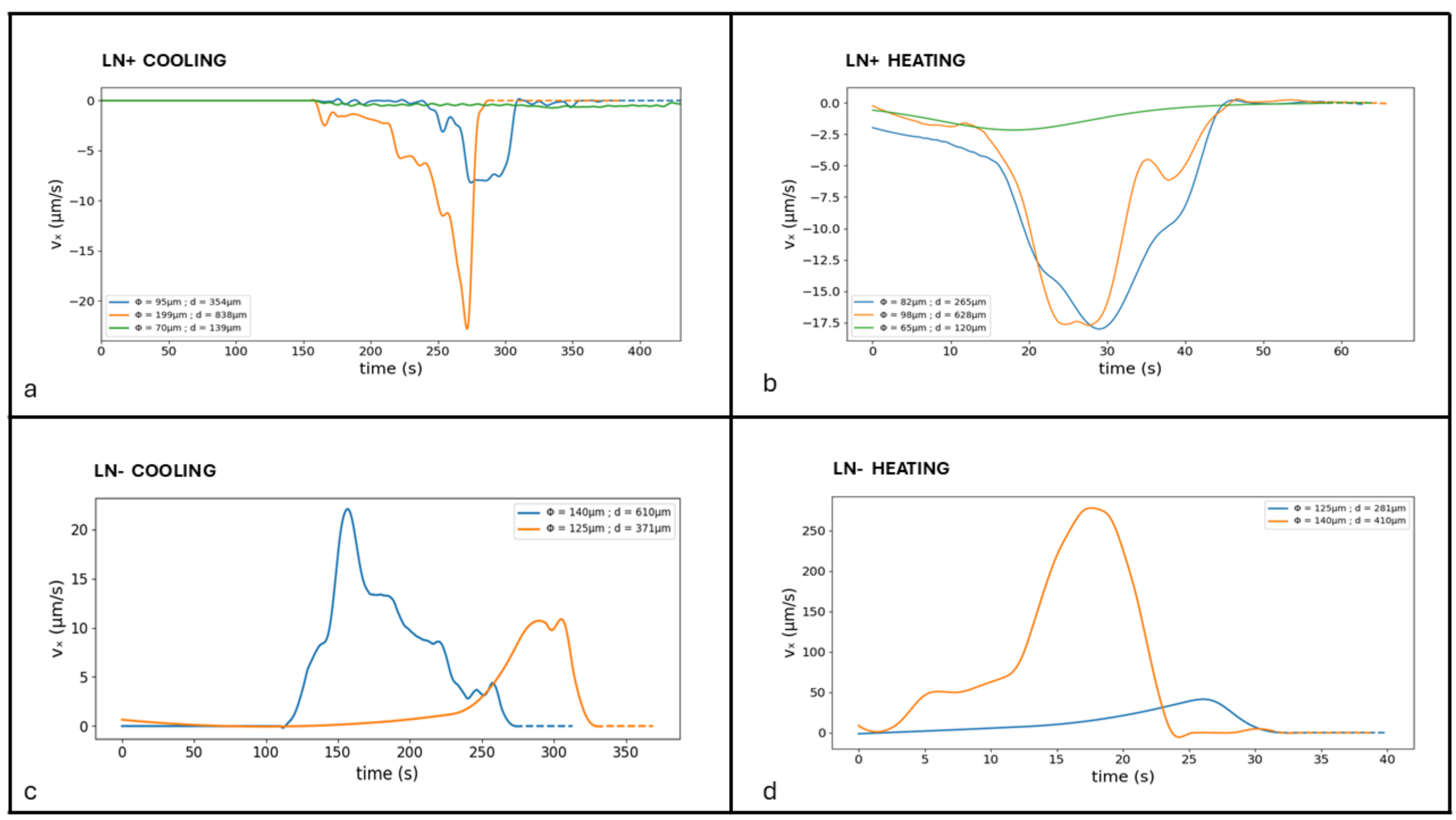

**Figure 5:** Droplets velocity as a function of time, obtained as a time derivative of Fig. 4. The speed depends on the single measurement. On average, it is higher during the heating cycles (b and d) and when the polar axis of LN crystals is negative (c and d).

With this approximation, the velocities in Fig. 5. gives us an approximate idea of the shape of the electrostatic force acting on the droplets. The figure confirms that the force profile is different for different droplets. It can be very smooth, as in the case of the smallest droplet in Fig. 5a, or quite steep as for the largest droplets in the same figure or in Figures 5c and d. On average, the force acting on $N_F$ droplets is stronger upon heating than upon cooling, due to the larger fringing field.
Moreover, the force is larger when LN crystals with negative polar axis are used, suggesting some asymmetry either in the value of the fringing field generated by crystals with opposite polar axes, or related to droplets electrification. Indeed, LN crystals with opposite polar axes generate fringing fields of opposite signs, which in turn affects the sign of the bound charge accumulated in the splayed area. Since contact electrification involves electron transfer to the substrate, we expect this process to be somehow dependent on the field sign. Although we could not find a final explanation for this behavior, we might tentatively assume that when the bound charge accumulated in the splayed area is negative, electrons are more easily released compared to what happens when this charge is positive. In this latter case electron release might be less likely, which would explain why the probability of observing droplets motion is higher in case of LN with positive polar axis. Interestingly, when using LN crystals with negative polar axis, it is common to observe the release of tiny pieces of fluid during the first seconds of motion upon cooling the system (see Video S7), an event not observed in the case of crystals with positive polar axis. This could be a further attempt to decrease the electrostatic energy by leaving behind small parts carrying an excess of negative charge. This process, which might be favored whenever contact electrification is not efficient enough, is typically followed by a droplet acceleration.

An example of how droplets' velocity depends on position is shown in Fig. 6, where only the case of LN with positive polar axis is shown. The velocity, and according to our rough assumption, the force, increases as droplets move and reaches a maximum at a certain distance from the initial position, which can be closer or further to the charged crystal. For instance, in Fig. 6a, the force initially increases on increasing the distance from the LN crystal, where $E_f$ is instead decreasing. We understand this behavior as due to an increase of droplet charging during motion, which can be either due to friction-induced charging, already observed for water droplets[32,39], or to the so-called "slide electrification"[40]. This latter has also been reported for water and is defined as the charging occurring each time a sliding droplet leaves some of its charge on the dewetted substrate surface. A redistribution of the ions in the $N_F$ droplet due to electrostatic interactions with the charged solid surface might also come into play[31].

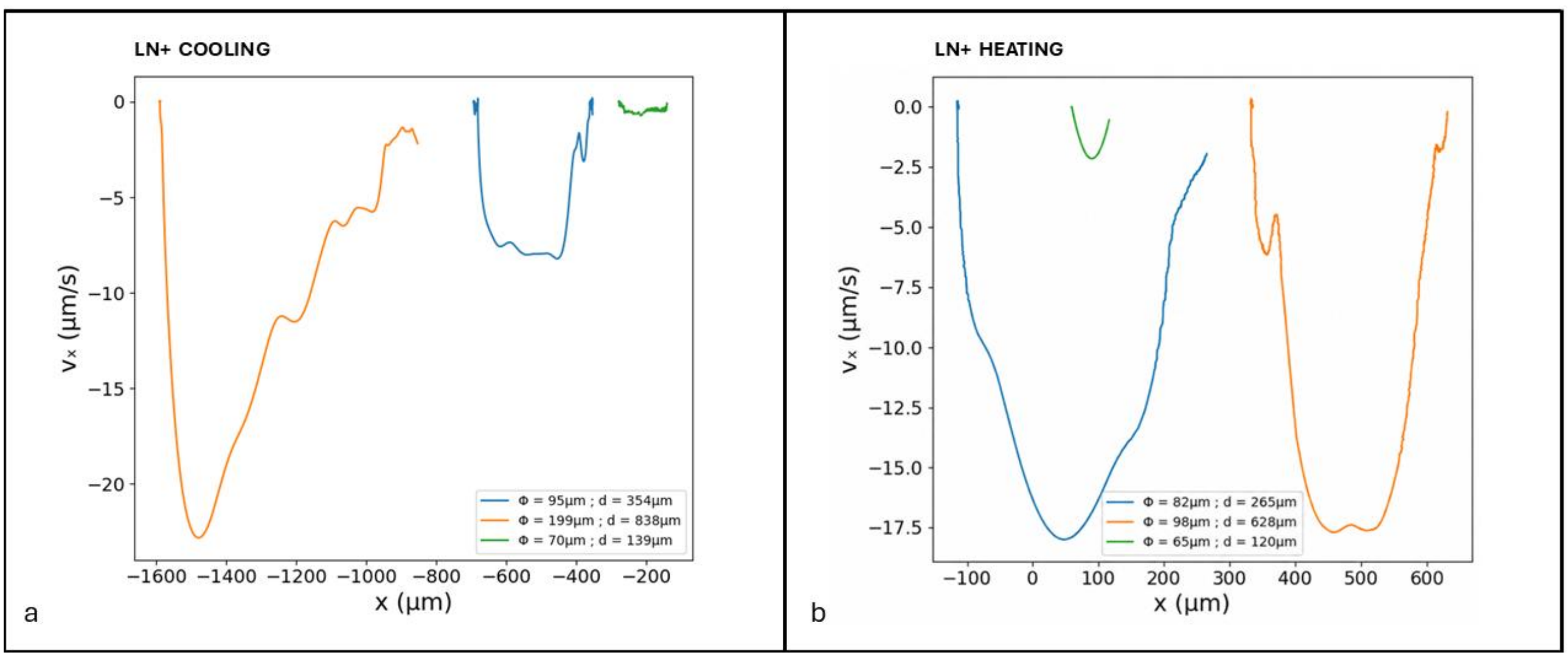

**Figure 6:** Droplets velocity as a function of position, obtained combining Figure 4 and Figure 5. Only the case of LN with positive polar axis is shown. The velocity varies as droplets move and reaches its maximum at a certain distance that can be more or less close to the edge of the LN. In (a) the velocity, and thus the force, increases as droplets move away from the crystal, in regions where the fringing field is decreasing, indicating that droplets charging become larger during motion (see text).

Additional experimental evidence in favor of the proposed splay-induced charging, is given by the (rare) observation of dielectrophoretic interaction in the first stages of motion, followed by electrophoretic-like response. An example of this is reported in Fig. 7, where the displacement of an $N_F$ droplet close to a LN crystal with positive axis, is shown during a cooling cycle. The droplet starts moving toward the border of the LN crystal, as one would expect in case of dielectrophoresis, while the repulsive motion comes into play just after tens of seconds. This clearly demonstrates that $N_F$ droplets become electrically charged after a certain time of exposition to the non-uniform fringing field. Before charging, the interaction with the field is governed by dielectrophoresis; the reason why this is observed only rarely indicates that dielectrophoretic force is most of the time too weak to overcome surface pinning.

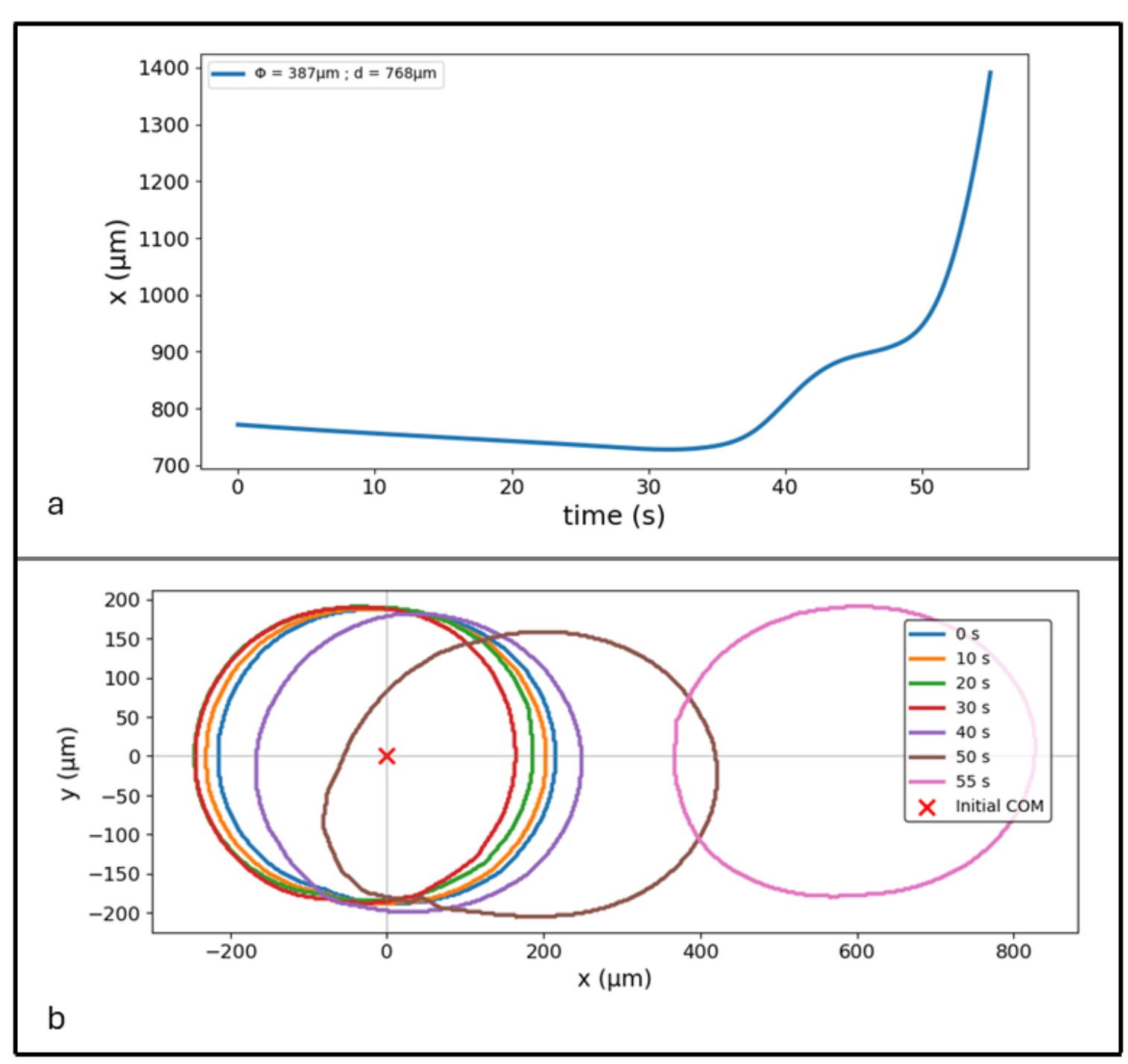

**Figure 7**: a) Droplet distance from the border of a LN crystal (x = 0) as a function of time, showing the droplet approaching the crystal during the first 33 s and then being repelled. t = 0 here marks the beginning of motion. The LN crystal has positive polar axis and the system is undergoing a cooling cycle; b) Droplet profiles at different times drawn as a function of the position of the droplet center, showing the different directions of motion. x = 0 marks here the initial position of the center of mass.

All the observations described above support the interpretation based on $N_F$ droplets charging. The delay observed between the transition to the $N_F$ phase on cooling and the beginning of motion is related to the time it takes to the droplets to generate regions of nonvanishing **P** divergence, and to develop electrostatic charge by contact electrification. The different speed indicates that the amount of charge is different from case to case, being related to droplet size, initial distance from the field source and other possible parameters that are not easily quantifiable. Overall, situations where droplets are large and are positioned in regions where the field non-uniformity is relevant on the scale of their dimensions, are the most favored. Notably, not all the droplets move, even in the same experiments. There are cases where more than one droplet is present and just one or two of them move (not necessarily those closer to LN, see Video S3-2), indicating that charging depends on the specific kind and probably amount of polarization distortion.

When splay-induced charging does not occur, droplets behave as neutral, polar objects. They move because of dielectrophoresis, very rarely as we saw, or they undergo deformation and eventually electromechanical instability with the emission of fluid jets. These latter effects are well highlighted using two LN crystals (see SI, Note 8). In these experiments, droplets are positioned on the PTFE coated coverslip, in correspondence of the gap between the crystals, at different distances from their borders, as shown in Fig. 1 b, b' and c, c'. The two crystals have either parallel or antiparallel polar axes, so that the surfaces underneath the PTFE-coated coverslip have either equal or opposite charge. The total fringing field in the two situations is different, being strongly non-uniform in case of parallel polar axes and quasi uniform when polar axes are antiparallel (Fig. 1 b'',b''' and c'',c''', respectively).

This difference translates into a different response of $N_F$ droplets.

When LN crystals have parallel axes, droplets positioned close to one of the two crystals move following the same rules as for the single LN, being repelled or attracted from the closest LN slab, depending on the sign of the polar axes and of the temperature gradient. When motion is not observed and electromechanical instability occurs, jets are emitted from regions in front of the two crystals and are asymmetrically distributed (see Fig. S9). This indicates that the tips of the jets are oppositely charged on the two sides, meaning that droplets are strongly polarized by the closest LN slab. As soon as jets separate from the mother droplets, they become free to move over the whole area, suggesting that the tips neutralize due to the redistribution of polarization charges (Fig. S9f). If droplets are instead positioned right in the center of the gap, where the in-plane component of the field vanishes, we do not observe any motion in any of our tests. Droplet fission sometimes occurs, being characterized by the emission of symmetrically distributed fluid jets travelling over both crystals.

When LN polar axes are antiparallel, droplet motion is never observed, regardless of droplet position with respect to the center of the gap, confirming that electrostatic charging of $N_F$ droplets requires highly non-uniform electric fields. Droplets exposed to this quasi-uniform electric field undergo significant elastic deformation, the extent of which depends on the electric field squared, and that sometimes ends up in droplets fission (SI, Note 8 and Fig. S10-11). We interpret this effect as an electrostriction-like mechanical deformation due to electrical polarization[30,41].

The experiments with two LN crystals further highlight that $N_F$ sessile droplets in a non-uniform electric field can behave either as charged particles or as neutral polar objects depending on how non-uniform the electric field is and on experimental parameters as droplet size and position and freshness of PTFE coating.

The total force *F* acting on a $N_F$ droplet can thus be written as the sum of several terms:

$$\boldsymbol{F} = \rho_{free}\boldsymbol{E} + (\boldsymbol{P}\cdot\boldsymbol{\nabla})\boldsymbol{E} + \frac{1}{2}\boldsymbol{\nabla}\left(E^2\rho_m\frac{\partial\varepsilon}{\partial\rho_m}\right) + \boldsymbol{f}_p + \boldsymbol{f}_v$$

where $\rho_{free}$ is the free charge density, **E** is the total electric field given by the LN fringing field plus the field generated by droplet polarization, $\rho_m$ is the RM734 mass density, $\varepsilon$ is its dielectric constant, $\mathbf{f}_p$ is the pinning force and $\mathbf{f}_v$ is the viscous friction.
The first three terms compete in defining the droplets' behavior. The first one dominates whenever unbalanced free charges are present. Indeed, dielectrophoresis, described by the second term and related to droplets polarization, is a mild effect compared to the motion of charged objects in an electric field[30], which thus prevails whenever $\rho_{free}$ is different from zero.
When the distortion of the polarization vector produced by the fringing field does not induce droplets charging, the first term in eq. (1) becomes negligible and the effects of polarization prevail. In our experiments this happens approximately 70% of the time in configurations with a single LN crystal and with two LN crystals with parallel polar axes. In these cases, droplets motion is not observed. This means that splay-induced droplets charging is not the most favored effect and that even if PTFE coating guarantees low pinning, dielectrophoresis is easily hindered by pinning forces.
The third term[42] relates to mechanical deformations of the droplet and becomes dominant in the case of quasi-uniform electric field, where splay induced charging cannot occur and dieletrophoretic motion is not favored. This term arises when surface forces due to pressure gradients are considered. Since real liquids are compressible, the pressure changes due to electrostatic energy are accompanied by changes of density and therefore of dielectric constant[30].

**Conclusions**

We reported on the behavior of $N_F$ droplets in the electric field generated by pyroelectrically charged LN crystals. We show that upon entering the ferroelectric nematic phase, RM734 droplets on PTFE-coated surfaces can acquire a net electric charge due to contact electrification and that this effect is favored when the field is highly non-uniform.

Our results show that droplets can act either as neutral polar media or as charged particles, depending on the specific situation. When the proper combination of droplets size, initial distance from the LN crystal, inhomogeneity of the fringing field and other experimental parameters is met, localized distortions of **P** give rise to splay-induced charging of the droplets, which thus respond to the field as charged objects. When charging does not occur, the interaction with the non-uniform field is of dielectrophoretic kind but is hindered by pinning most of the time. Finally, when the field inhomogeneity is too weak to induce either localized splay or dielectrophoretic motion, a large deformation dependent on the electric field squared is observed. We understand this deformation as an electrostriction-like effect due to droplet polarization.

The described splay-induced charging emerges as a new property of ferroelectric nematics when confined in form of sessile droplets. Noteworthy, self-charging of isolated $N_F$ droplets embedded in their isotropic phase has been very recently observed[43]. We expect that, upon optimizing the minimization of surface pinning, the reported mechanisms might be tuned acting on the ratio between droplets size and field inhomogeneity to obtain either electrophoretic or dielectrophoretic motion, which might be desirable in microfluidic applications. As an additional remark, splay-induced contact electrification of $N_F$ droplets, might find applications in the development of triboelectric nanogenerators (TENG) based on ferroelectric liquids/PTFE systems for energy harvesting.

The electromechanical responses we observe whenever droplets charging does not occur, namely instability and nonlinear mechanical deformations, are also peculiar to the $N_F$ phase. Understanding the electromechanical response of this phase of matter appears desirable for the development of fluid actuators based on the unique combination of polarization and fluidity.

**Acknowledgements**

Authors acknowledge Peter Salamon for suggesting the PTFE coating and providing the proper receipt and Tommaso Bellini and Pierangelo Metrangolo for useful discussion.

This research was sponsored by the Army Research Laboratory and was accomplished under Grant Number W911NF-24-1-0252. The views and conclusions contained in this document are those of the authors and should not be interpreted as representing the official policies, either expressed or implied, of the Army Research Laboratory or the U.S. Government. The U.S. Government is authorized to reproduce and distribute reprints for Government purposes notwithstanding any copyright notation herein.

**Conflict of interest**

Authors declare no conflict of interest

Supplementary Information

**Splay-induced charging of ferroelectric droplets in highly non-uniform electric fields**

Lorenzo Fiorentini, Elia Rocchetti, Raouf Barboza, Roberta Galeazzi, Liana Lucchetti*

1. *Dipartimento SIMAU, Università Politecnica delle Marche, via Brecce Bianche, 60131 Ancona, Italy. E-mail: l.lucchetti@univpm.it*

2. *Dipartimento DISVA, Università Politecnica delle Marche, via Brecce Bianche, 60131 Ancona, Italy.*

**Note 1: preparation of RM734 droplets**

RM734 droplets were obtained following two steps. First, a small amount of RM734 powder is deposited at room temperature on a clean glass slide and heated to 150°C to have a melt. To create the initial droplets, a cold stainless needle is dipped into the melt and retracted, so that the droplet on its tip solidifies at contact with the surrounding air. Then, the pearl is remolten into a droplet on the glass substrate. The glass slide is then cooled down to room temperature so that solidified droplets can be peeled off and reused on the proper substrate. The sizes of the droplets are controllable (size before peeling vs size of the remolten droplet) and are measured by means of a calibration slide.

**Note 2: PTFE coating**

For the realization of the PTFE-coated substrates, we prepare a 1 wt% solution of the amorphous fluoropolymer poly[4,5-difluoro-2,2-bis(trifluoromethyl)-1,3-dioxole-co-tetrafluoroethylene] in FC-40. Dissolution of the polymer in FC-40 is promoted through stirring for at least 2 days; after two days of stirring, the dissolution process can be accelerated by heating the solution to 70–80 °C while keeping the container closed to prevent solvent evaporation. For each substrate, 30 μL of the solution are deposited onto 1 cm × 1 cm × 100 μm glass coverslip and then spin coated to obtain a layer of uniform thickness. The spin-coating program consists of an initial 10 s acceleration from 0 to 2000 rpm, followed by 1 min at 2000 rpm, and a final 10 s deceleration from 2000 rpm to 0. After coating, samples are left to dry in air for 5 min and then thermally annealed for 5 min at 50 °C, 10 min at 245 °C, and finally 30 min at 300 °C to consolidate the fluoropolymer film on the glass substrate. Both the prepared solution and the coated substrates are used immediately, since a loss of effectiveness has been observed over the course of 24 hours.

**Note 3: Contact angle**

Measurements of the contact angle as a function of temperature were performed with the set-up shown in Fig. S1, made of a collimated light source, an imaging system, and a tiltable hot stage. A white LED, an iris and a collimating lens are used to produce the collimated light source, while the imaging system is made of a collecting and an imaging lens with their focal back-to-back, an iris and a CMOS camera. Such a configuration provides the needed tele-centricity and a constant magnification ratio. To control the LC temperature, we use a slotted aluminum plate able to host resistive cartridges powered by a programmable power supply. The temperature of the hot stage is sampled with a thermocouple connected to a DAQ. The desired set point and the measured temperature are sent to a Labview PID controller that drives the programmable power supply.

Droplets contact angles were measured in a tilted configuration (17°) to evaluate the difference between advancing and receding angles, which gives an indication on the strength of the pinning force[1] (see Fig. S2).

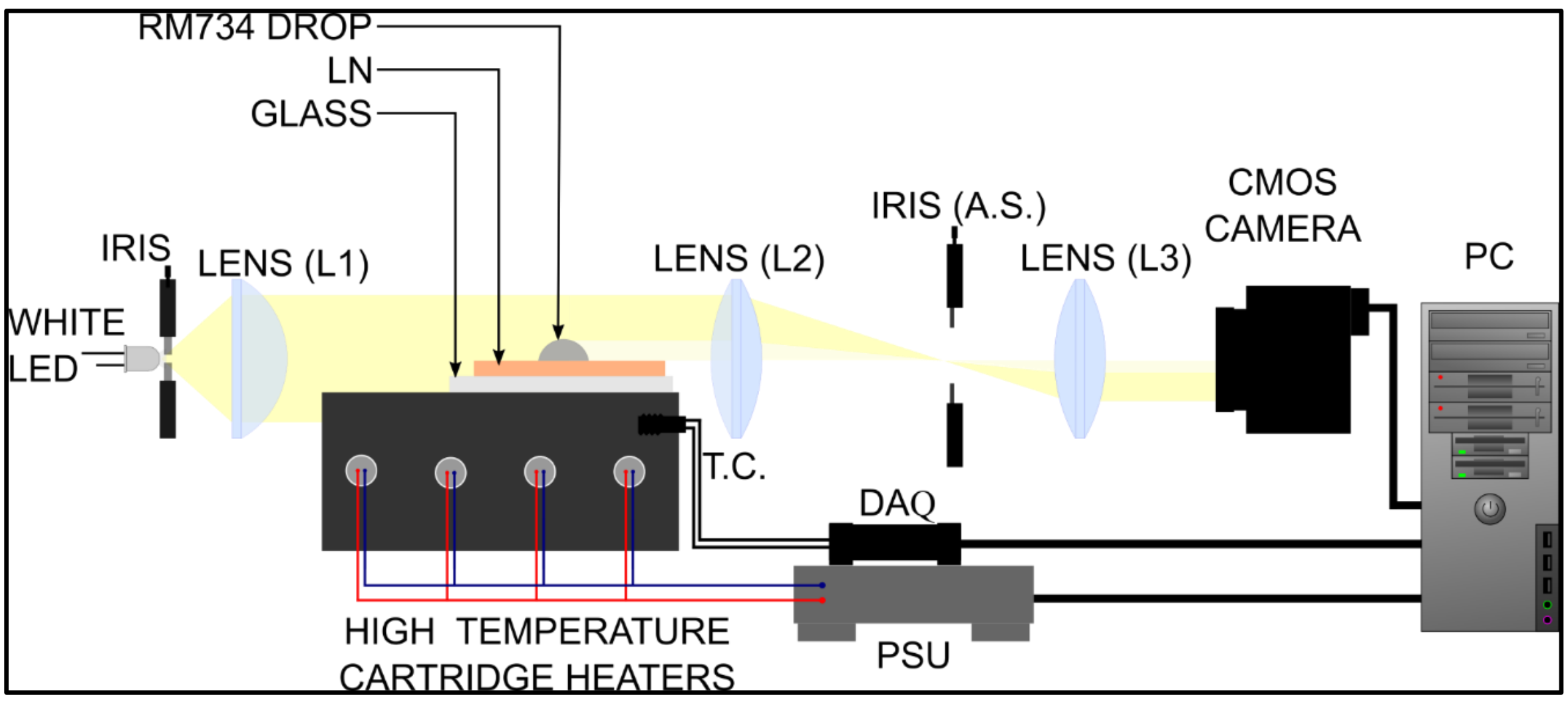

**Figure S1**: Set up for measuring the contact angle of RM734 droplets on LN substrates. L1 collimating lens. L2 and L3 imaging and collecting lens. The first iris is used to reduce the width of the LED, the second as an aperture stop (A.S.). The CMOS sensor is placed in the focal plane of L3. The temperature of the LN substrate is controlled by a heating stage made of a slotted aluminium plate fitted with resistive heater cartridges powered by a programmable power supply unit (PSU). The setting point is precisely controlled (within 1.5 °C) by feeding the measured temperature with a thermocouple (T.C.) and DAQ to a PID controller that drives the PSU.

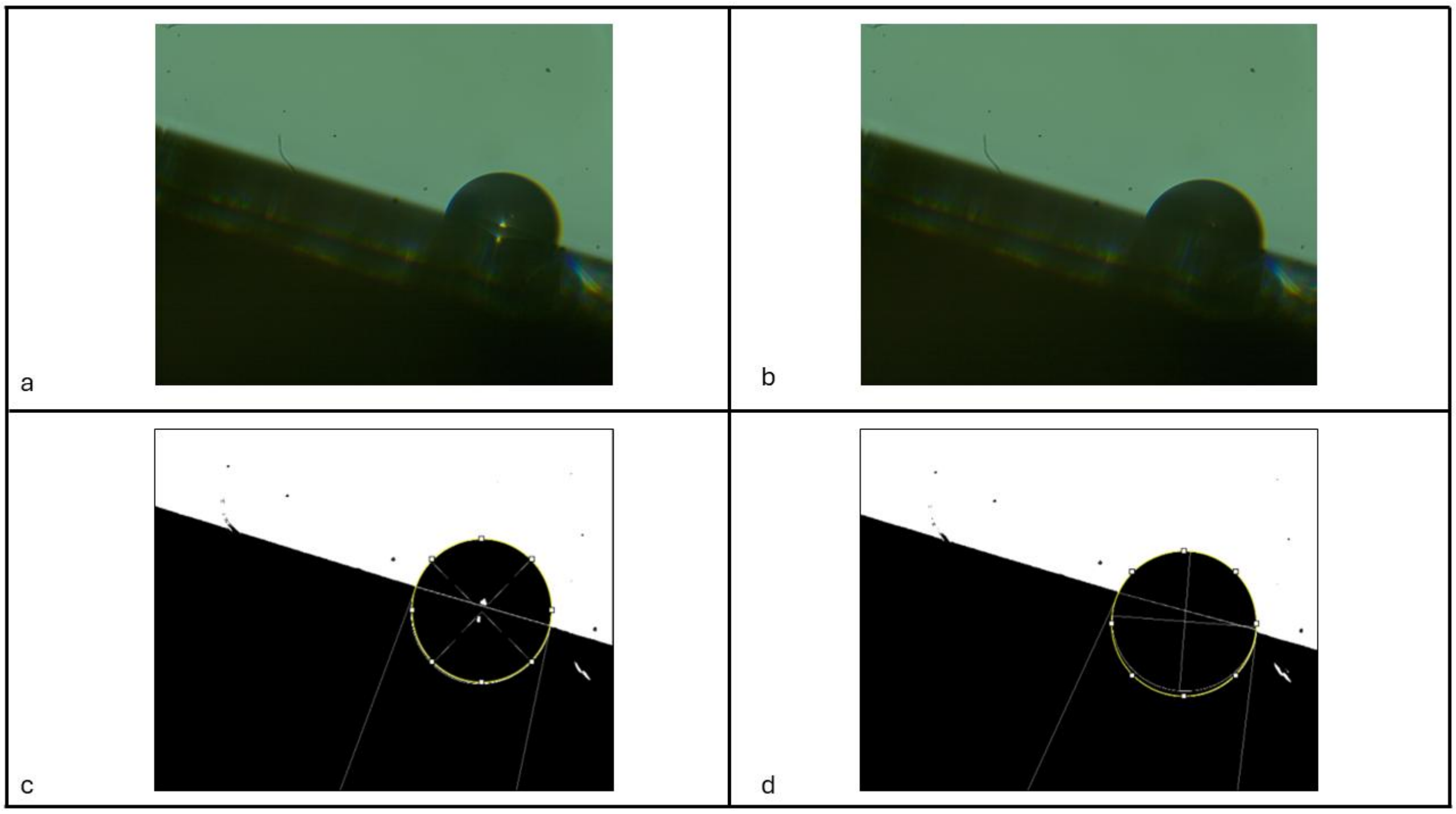

**Figure S2**: Contact angle of a RM734 sessile droplet on glass coated by PTFE. Pictures of droplet in the nematic (a) and ferroelectric nematic (b) phase. c,d) Binarized images used for the evaluation of the contact angles in the nematic (c) and ferroelectric nematic (d) phase. The values of advancing and receding angles extracted from the two images are: 99.8° and 98.2° in the nematic phase and 95.2° and 93.8° in the ferroelectric nematic phase.

**Note 4: RM734 droplets under a uniform electric field**

To investigate the behavior of RM734 $N_F$ droplets under an external uniform electric field, we first equipped a PTFE-coated coverslip with a custom electrode configuration. The two short sides of the coverslip were covered with copper, leaving a free gap of 1 mm between the two copper regions (see Fig. S3). Electrical wires were then soldered onto the copper coated areas and connected to a function generator through a 10× voltage amplifier, allowing the generation of a uniform electric field across the gap. The applied signal was supplied in both AC and DC mode and monitored using an oscilloscope. The amplifier enabled a maximum applied voltage of 500 V across the electrode gap. The central glass region between the copper electrodes, where the RM734 droplet was deposited, was coated with PTFE following the procedure described in Note 2. During the measurements, the entire structure was placed on a heating plate and maintained at 120 °C to ensure that droplet was in its ferroelectric nematic phase.

In this configuration, droplet motion was not observed.

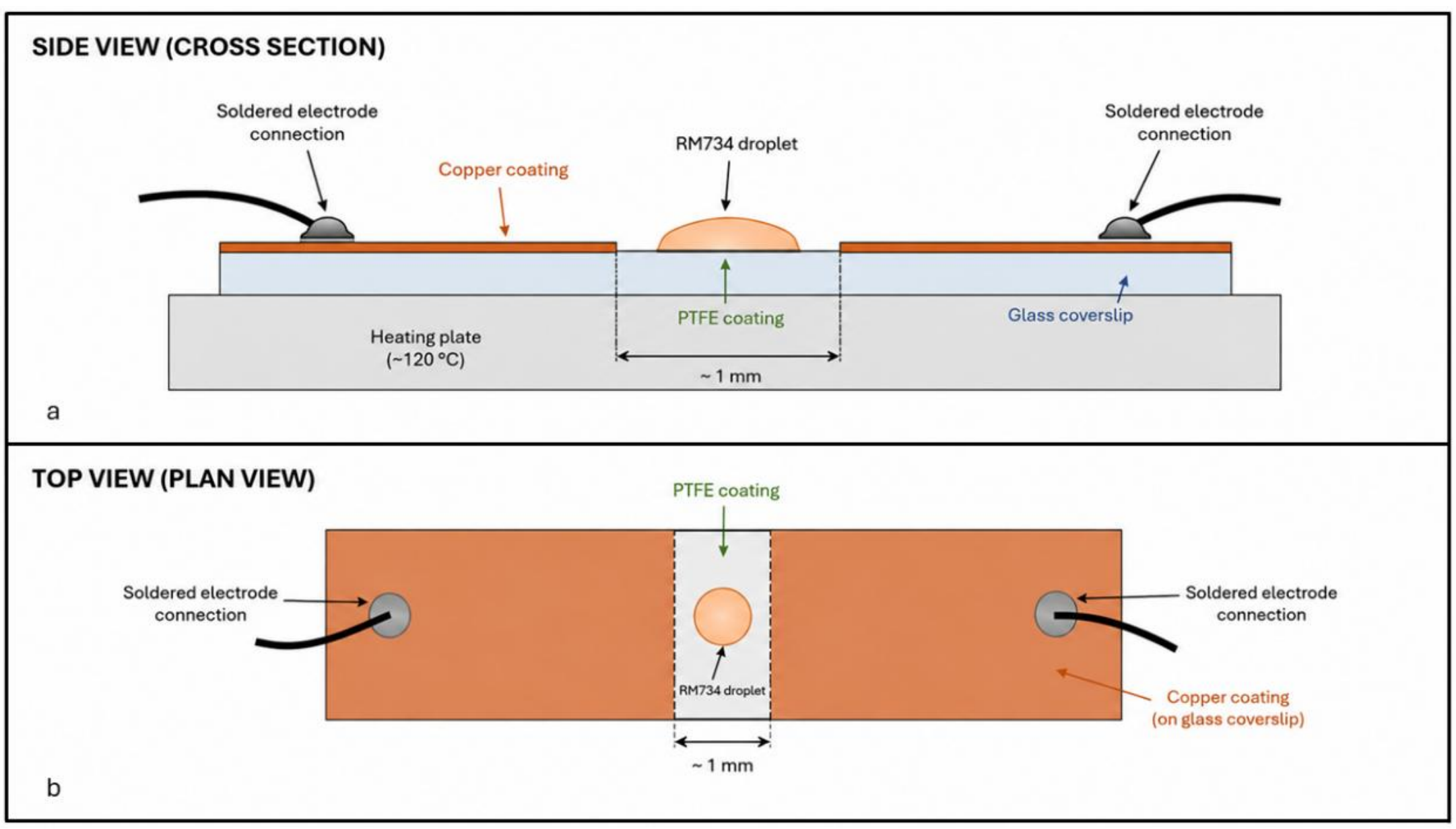

**Figure S3:** Side (a) and top (b) views of the experimental arrangement realized to study the behaviour of $N_F$ droplets under a uniform electric field. The field has been applied using copper electrodes and a function generator connected to a voltage amplifier.

One possible reason for the absence of motion of a charged sessile droplet in a uniform electric field, could be related to a too weak value of the field, which, combined with a possible weak charge, gives rise to a force that does not overcome surface pinning. To be sure to use an electric field comparable to the one at play in the experiments described in the main text, one of the z-cut LN crystal was employed as a generator of surface charge. To this purpose, both polar faces of the LN crystal were coated with copper. Electrical contacts were then applied without soldering them directly onto the crystal, so to avoid the formation of microcracks or mechanical damage within the LN crystal. Instead, the contacts were wrapped with copper and placed in intimate contact both with the crystal and the conductive coating with the aid of conductive paste, avoiding air gaps or cavities. A sketch of the experimental arrangement is shown in Fig. S4.

The assembled LN crystal-based generator was placed on a heating plate and heated from room temperature to 165 °C, allowing the pyroelectric field to develop. During the heating process, the two pyroelectrically charged LN surfaces were coupled to the external electrodes and used to generate a uniform field comparable to the fringing field generated by the same surface charge in the experiments of the main text. The droplet substrate was maintained on a hot stage at 120 °C to ensure that RM734 was in the ferroelectric nematic phase during the experiment.

The main difference between the configuration described here and the one used in the main text, resides in the profile of the electric field obtained. In the main text, the fringing field is due to the internal pyroelectric field combined with the finite size of the crystal. This field extends in the region where $N_F$ droplets are positioned and is strongly non uniform. On the contrary, in the experiment reported here, $N_F$ droplets are far enough from LN to be not affected by the residual fringing field. The field they feel is instead the one generated by the two electrodes that have been charged using the pyroelectric charges of the two LN surfaces, which are supposed to be equal in value and opposite in sign.

Again, no motion was observed.

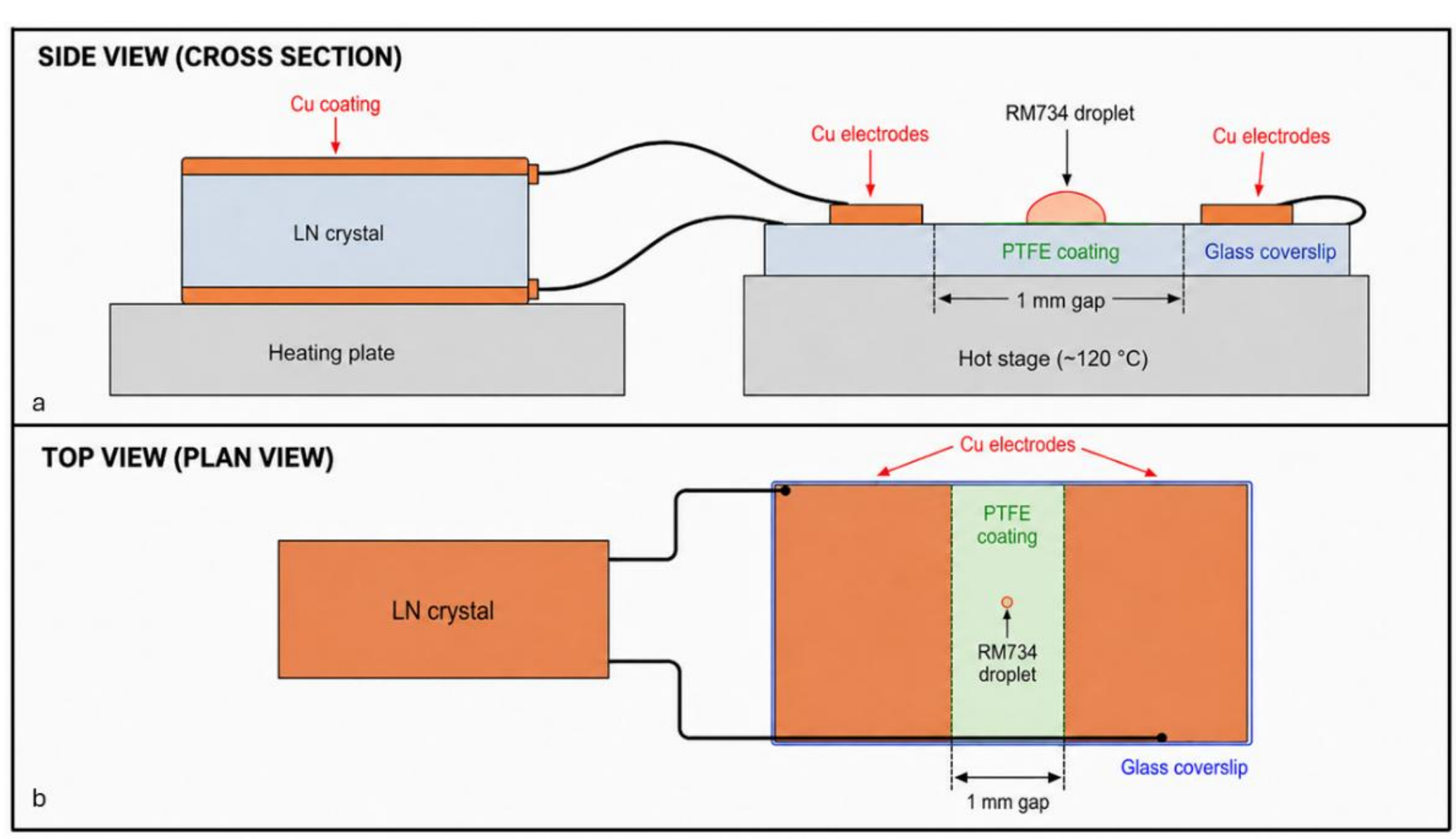

**Figure S4:** Side (a) and top (b) views of the experimental arrangement realized to study the behaviour of $N_F$ droplets under a uniform electric field applied using a z-cut LN crystal as a source of electrostatic charges.

### Note 5: Droplet motion: dielectrophoresis and possible apo-electric behavior

If $N_F$ droplets remain neutral, the non-uniform fringing field exerts on them a dielectrophoretic force[2]. This kind of force should always produce a motion toward the region of maximum electric field, i.e. the border of LN crystal, contrary to our observations.

Exceptions to this rule are possible in case of materials possessing spontaneous polarization[2]. Specifically, if the system is already strongly polarized (e.g., a permanent dipole) and is spinning[2] an apo-electric response can occur — one with a dielectrophoretic force directed oppositely to the region of strongest field[2]. In the case of RM734 droplets, Salomon et al. reported the onset of a circular flow of micrometric particles at the transition to the $N_F$ phase. However, this flow does not seem to be associated with a rotation of the molecular director, which is instead at rest[3].

Apo-electric behavior is well known in the case of molecules. In certain rotational states they are repelled by the strongest field regions, since these are the regions of highest energy[2]. The case of macroscopic objects is way less known.

A tentative simple interpretation in the case of RM734 droplets could be given based on energetic considerations. Neglecting the induced polarization term (which is several orders of magnitude lower than the spontaneous polarization, see Note 6), one might expect apo-electric behavior each time the polarization vector **P** is antiparallel to the electric field and paraelectric behavior (conventional attractive dielectrophoretic force) each time **P** is parallel to the electric field. In our experimental conditions, this would mean that in the case of LN crystals with positive axis upon cooling, where a repulsive force is observed, the LC polarization vector is always antiparallel to the field. The opposite behavior observed on heating would require that the **P** keeps its direction, thus being now parallel to the field, which is of course not reasonable. Indeed, the coupling to the unrubbed PTFE surface cannot be responsible for a predetermined preferential direction of **P** and even if it were, such preferred direction cannot be insensitive to field reversal. This is clearly demonstrated in experiments on thin sandwich cells prepared by confining an RM734 droplet between two PTFE coated glasses. These cells exhibit random planar alignment in both the nematic phases (Fig. S5 a-b'). When an LN crystal is put close to the cells and the system is cooled down, the polarization vector **P** in $N_F$ phase aligns on average in the direction of the field, as shown in Fig. S5 a'' and b'', and reverts its sign soon after one passes from cooling to heating (i.e. soon after the sign of the electric field changes) (see Video S5).
In addition, we highlight two fundamental points: i) since droplets roll over the PTFE-coated substrate, the average direction of **P** cannot remain constant during droplets motion; ii) droplets that move toward the LN crystal, typically do not stop in close proximity of the border but keep on moving over the crystal, suggesting that they are attracted by the charged surface rather than by the region of maximum field.
Droplets charging is thus the most reasonable interpretation of our results.

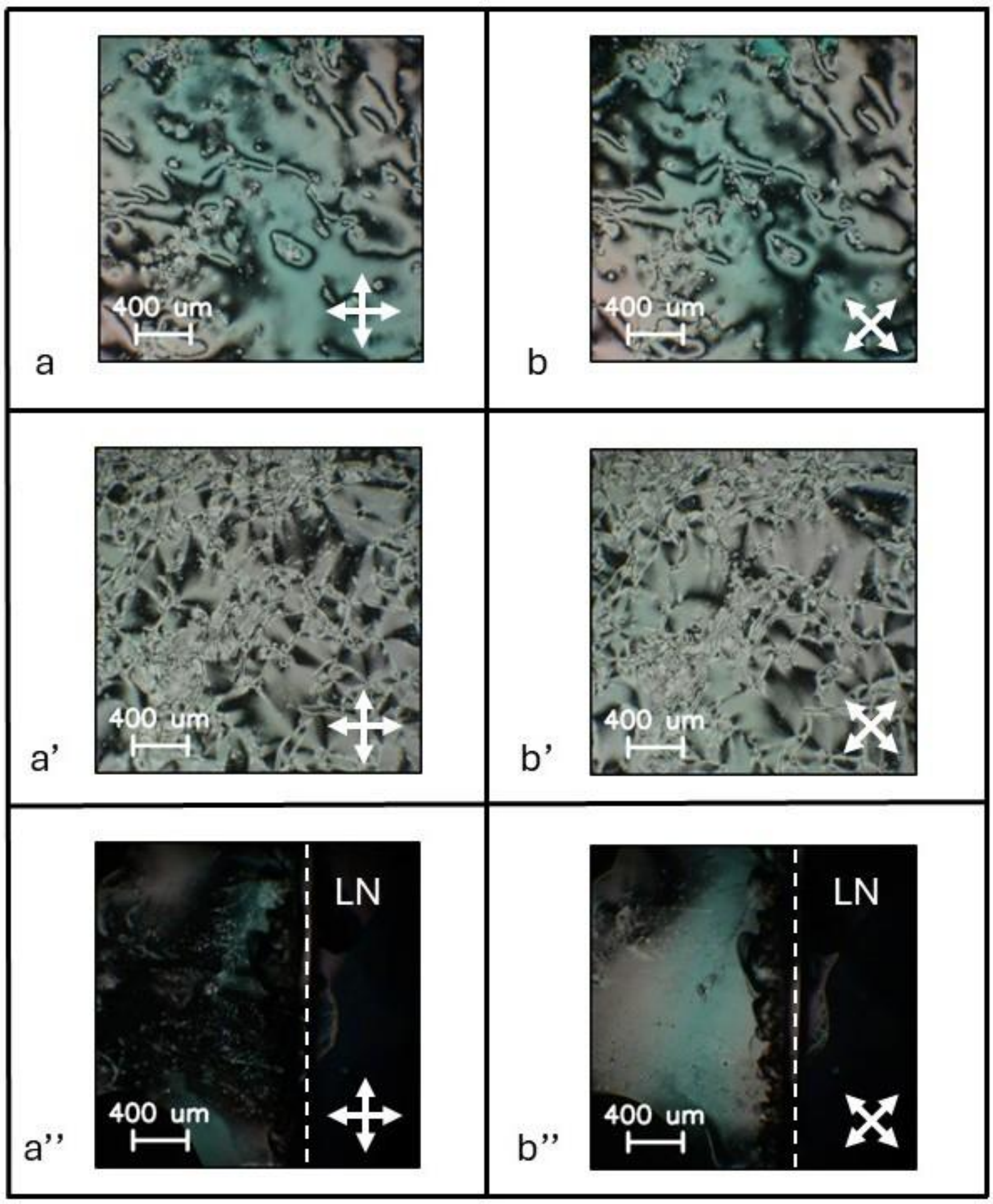

**Figure S5:** Thin LC cell (thickness d = 10 μm) realized confining an RM734 droplet between two PTFE-coated glass substrates. Cell appearance in the nematic phase (a,b) and in ferroelectric nematic phase (a',b'). The alignment is random planar in both phases. When a LN crystal is positioned close to the cell, the director **n** and the polarization vector **P** are affected by the fringing field and align along its direction (a'',b''). White dashed line marks the LN border. LC cell is positioned on the left of the line. White crosses represent the axes of the optical microscope polarizers.

## Note 6: Permanent vs induced polarization

To have an idea of the difference between spontaneous and induced polarization of a liquid crystal in the $N_F$ phase under an external electric field, we can compare the related energy densities[4].
The characteristic ferroelectric energy density in an external electric field E is of order $\frac{1}{2}PE$, while the characteristic dielectric energy density is of order $\frac{1}{2}\varepsilon_0\Delta\varepsilon E^2$, where Δε is the $N_F$ dielectric anisotropy. Assuming P = 5 × $10^{-2}$ C/m$^2$, which is the value at T = 120°C, the temperature that typically corresponds to the beginning of droplets motion, and E = $10^5$ V/m (see main text), with Δε = 10 we obtain the following values: about 2.5 x $10^3$ J/m$^3$ for the ferroelectric coupling and about 4.5 x $10^{-1}$ J/m$^3$ for the dielectric term. Therefore, the characteristic dielectric energy density is about 4 orders of magnitude smaller than the ferroelectric energy density.

**Note 7: Density Functional Theory Analysis of RM734**

When a RM734 droplet is deposited on a PTFE-coated surface, an electron can be released from the highest occupied molecular orbitals (HOMO) of the RM734 molecule to the lowest unoccupied molecular orbitals (LUMO) of PTFE. A parameter that can be used to verify this process is the work function of the two materials, as reported in the main text. The work function of PTFE has been calculated in ref. [36] of the main text, while that of RM734 has been evaluated by means of DFT as detailed below, where it is indicated as Vertical Ionization Potential (VIP). According to the IUPAC Golden book, VIP is defined as the energy needed to remove an electron from an atom or molecule without changing its nuclear shape.

*Computational Methods*

All electronic-structure calculations were performed with Gaussian 16 Rev. B.01[5]. Geometry optimization was carried out at the B3LYP-D3(BJ)/6-311++G(d,p) level using tight SCF convergence and an ultrafine integration grid, followed by harmonic frequency analysis at the same level[6-8]. The optimized structure was then used without further relaxation for the ωB97X-D/6-311++G(d,p) neutral single-point, NBO/polarizability, TD-DFT, cationic, and anionic calculations. Twenty singlet excited states were calculated using linear-response TD-DFT at the ωB97X-D/6-311++G(d,p) level[9].
Vertical ionization potential (VIP) and vertical electron affinity (VEA) were evaluated by the ΔSCF method from single-point energies of the neutral, doublet cation, and doublet anion at the neutral optimized geometry[10-12]. Conceptual DFT descriptors were derived from VIP and VEA[10-11]. Unless otherwise stated, energies are reported in Hartree (Ha) or electronvolts (eV), using 1 Ha = 27.2114 eV. All supplied Gaussian jobs terminated normally.

*Geometry optimization and vibrational analysis*

The B3LYP-D3(BJ)/6-311++G(d,p) geometry optimization converged successfully to a stationary point with a final electronic energy of -1505.54071814 Ha. Harmonic frequency analysis yielded 138 real vibrational modes and no imaginary frequencies; the lowest calculated frequency was 8.8554 $cm^{-1}$. The zero-point vibrational correction was 0.356313 Ha, and the sum of electronic and thermal free energies was -1505.248640 Ha. These results confirm that the optimized gas-phase structure corresponds to a local minimum on the potential-energy surface.

**Table 1**. DFT Geometry-optimization and harmonic-frequency results obtained at B3LYP-D3(BJ)/6-311++G(d,p) level of theory.

| **Optimization/frequency quantity** | **Value** |
|---|---|
| Final optimized electronic energy | -1505.54071814 Ha |
| Number of vibrational modes | 138 |
| Imaginary frequencies | 0 |
| Lowest harmonic frequency | 8.8554 $cm^{-1}$ |
| Zero-point correction | 0.356313 Ha |
| Sum of electronic and thermal free energies | -1505.248640 Ha |
| Job status | Normal termination |

*Electronic Properties*

The final electronic energy obtained from the ωB97X-D/6-311++G(d,p) single-point calculation for the neutral molecule was -1504.92148941 Ha. The restricted closed-shell wavefunction contains 220 electrons (110 doubly occupied molecular orbitals), and the calculation terminated normally.
The calculated **dipole moment** is 11.516 D. Its Cartesian components are $\mu x$ = -11.0319 D, $\mu y$ = -3.1563 D, and $\mu z$ = -0.9768 D. Thus, the dipole vector is predominantly oriented along the input x direction, showing that the permanent dipole is strongly directional. Assignment of this axis to the molecular long axis should be confirmed visually from the optimized structure.
The calculated value differs from the previously reported value of 11.4 D[13] by approximately 1.0%.

**Table 2. Summary of calculated electronic properties for the optimized geometry**

| Property | Value |
|---|---|
| Electronic energy | -1504.92148941 Ha |
| Number of electrons | 220 (closed shell; 110 occupied MOs) |
| Dipole moment | 11.516 D |
| Previously reported dipole moment | 11.40 D |
| Isotropic polarizability | 313.691 $Bohr^3$ |

The HOMO and LUMO energies are -9.65 and -0.89 eV, respectively, yielding a HOMO-LUMO gap of 8.75 eV. Within the adopted Kohn–Sham orbital description, the relatively large HOMO–LUMO gap is consistent with a stable closed-shell electronic configuration and limited propensity for low-energy ground-state electron redistribution.

**Table 3. Frontier molecular-orbital energies**

| Orbital | Energy (Ha) | Energy (eV) |
|---|---|---|
| HOMO | -0.35446 | -9.645 |
| LUMO | -0.03286 | -0.894 |
| HOMO-LUMO gap | 0.32160 | 8.752 |

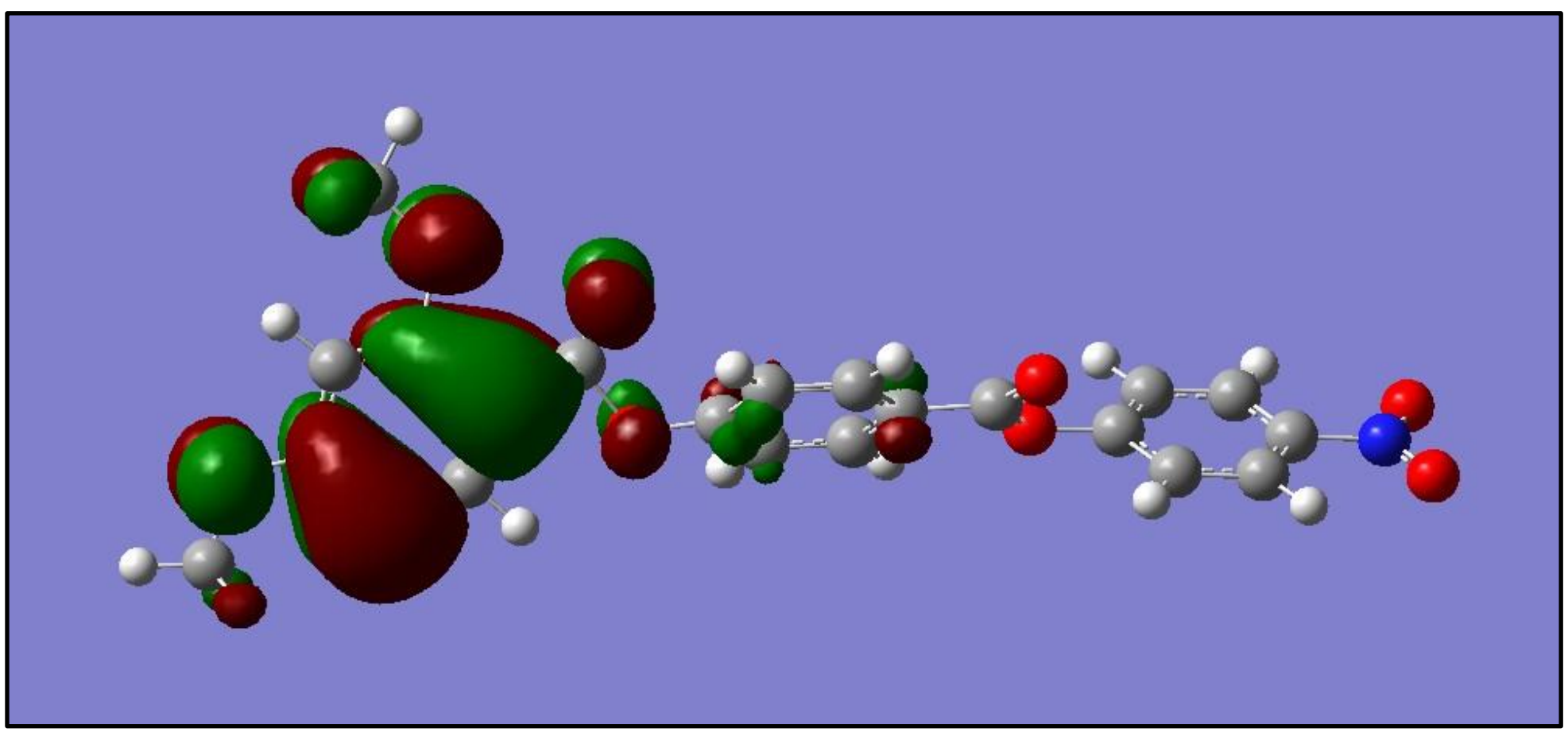

**Figure S6.** Isosurface representation of the HOMO (MO 110), plotted at an isovalue of 0.02 a.u.

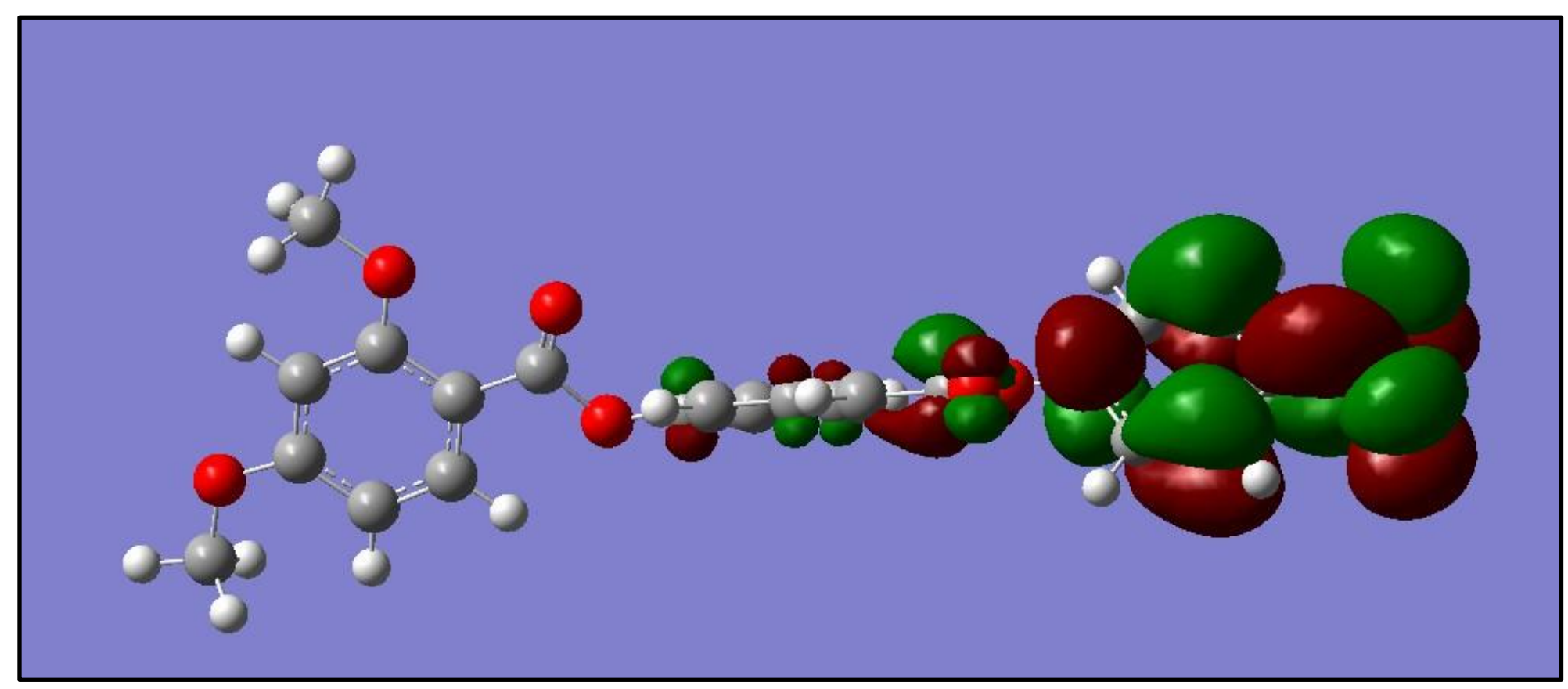

**Figure S7.** Isosurface representation of the LUMO (MO 111), plotted at an isovalue of 0.02 a.u.

*ΔSCF Ionization, electronic properties and DFT descriptors*

The ΔSCF values are preferred over simple Koopmans-type estimates[14] because they include electronic relaxation through separate total-energy calculations for the N, N-1, and N+1 electron systems. The cation and anion were evaluated vertically at the neutral optimized geometry. Results are reported in Table 4.

**Table 4. Summary of ΔSCF results**

| Quantity | Value |
|---|---|
| Neutral energy, E(N) | -1504.92148941 Ha |
| Cation energy, E(N-1) | -1504.61020433 Ha |
| Anion energy, E(N+1) | -1504.95210887 Ha |
| Vertical ionization potential, VIP | 8.47 eV |
| Vertical electron affinity, VEA | 0.83 eV |
| Electronegativity, χ | 4.65 eV |
| Chemical potential, μ | -4.65 eV |
| Chemical hardness, η | 3.82 eV |
| Softness, S | 0.131 $eV^{-1}$ |
| Electrophilicity index, ω | 2.83 eV |

The electronic descriptors are calculated according to the following equations:

$$VIP = E(N - 1) - E(N)$$
$$VEA = E(N) - E(N + 1)$$
$$\chi = (VIP + VEA) / 2$$
$$\mu = -\chi$$
$$\eta = (VIP - VEA) / 2$$
$$S = 1 / \eta$$
$$\omega = \mu^2 / (2\eta)$$

For comparison, Koopmans-type estimates derived from the frontier orbital energies give an ionization potential of 9.65 eV and an electron affinity of 0.89 eV, whereas the ΔSCF calculations yield a Vertical Ionization Potential of 8.47 eV and a vertical electron affinity of 0.83 eV. The close agreement for the electron affinity contrasts with the larger difference in the ionization energy, reflecting the fact that the ΔSCF approach includes electronic relaxation in the cationic and anionic states and is therefore expected to provide a more reliable description than the frozen-orbital Koopmans approximation. The obtained VIP is lower than the corresponding quantity calculated for PTFE (see main text), indicating that contact electrification at the interface between $N_F$ droplets and PTFE involves the release of electrons from the ferroelectric fluid to the superhydrophobic surface.

*Molecular Electrostatic Potential and Charge Distribution*

For completeness, we report the molecular electrostatic potential mapped onto the electron-density surface. It reveals the most negative-potential regions around the nitro and carbonyl oxygen atoms. More positive regions are localized predominantly near the hydrogen-bearing terminal portions of the molecule. The resulting electrostatic asymmetry is consistent with the large calculated permanent dipole moment.

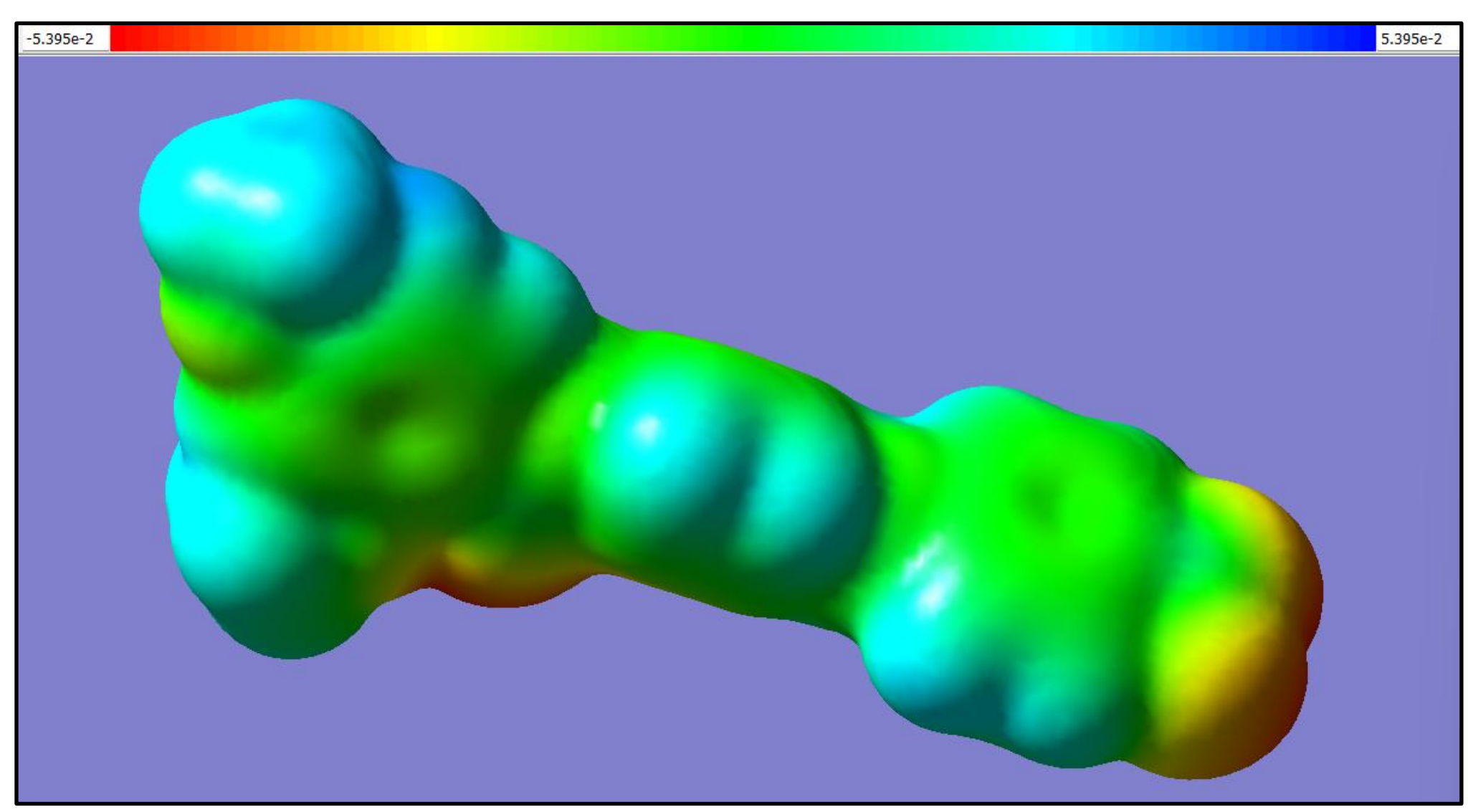

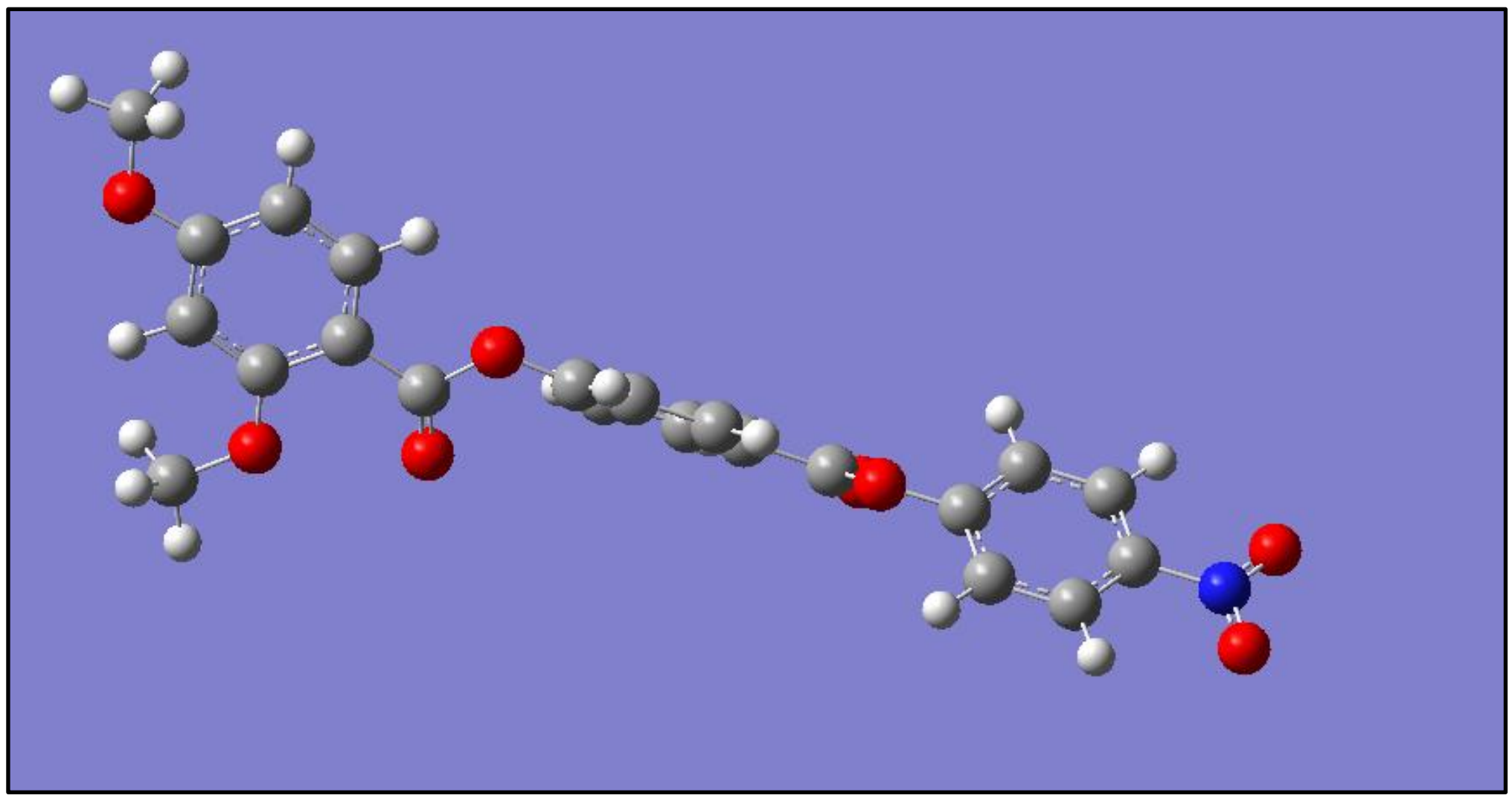

**Figure S8.** Upper panel: molecular electrostatic potential mapped onto the 0.004 a.u. electron-density isosurface. The potential scale ranges from $-5.395\times10^{-2}$ to $+5.395\times10^{-2}$ a.u.; red and blue indicate negative and positive electrostatic potential, respectively. Lower panel: ball-and-stick representation of RM734 in the same molecular orientation.

**Note 8: Experiments with two LN crystals**

*Parallel polar axes*

When two LN crystals with parallel polar axes are used, motion can be observed only if droplets are not positioned at the gap center. When they are closer to one of the two borders and move, they follow the same rules observed in the case of the single LN crystal. This means that they are attracted or repelled by the closest crystal, depending on the sign of its polar axis and of the temperature variation. When charging does not occur, droplets that are positioned closer to one of the two crystals, as in Fig. S9 a, typically undergo deformation followed by electromechanical instability with the ejection of fluid jets that travel over the closest crystal but are repelled by the further one. Only when jets detach from the droplet, they become free to move across this region. This is shown in Fig. S9, in the case of a heating cycle. The behavior described indicates that the tips of the jets are oppositely charged on the two sides, meaning that droplets are strongly polarized under the field of the closest LN slab. As soon as jets separate from the mother droplets, their tips neutralize due to the redistribution of polarization charges. Instability and jets ejection can be observed also when $N_F$ droplets are positioned at the center of the gap between the two crystals. However, in this case droplet fission is characterized by the emission of symmetrically distributed fluid jets travelling over both crystals.

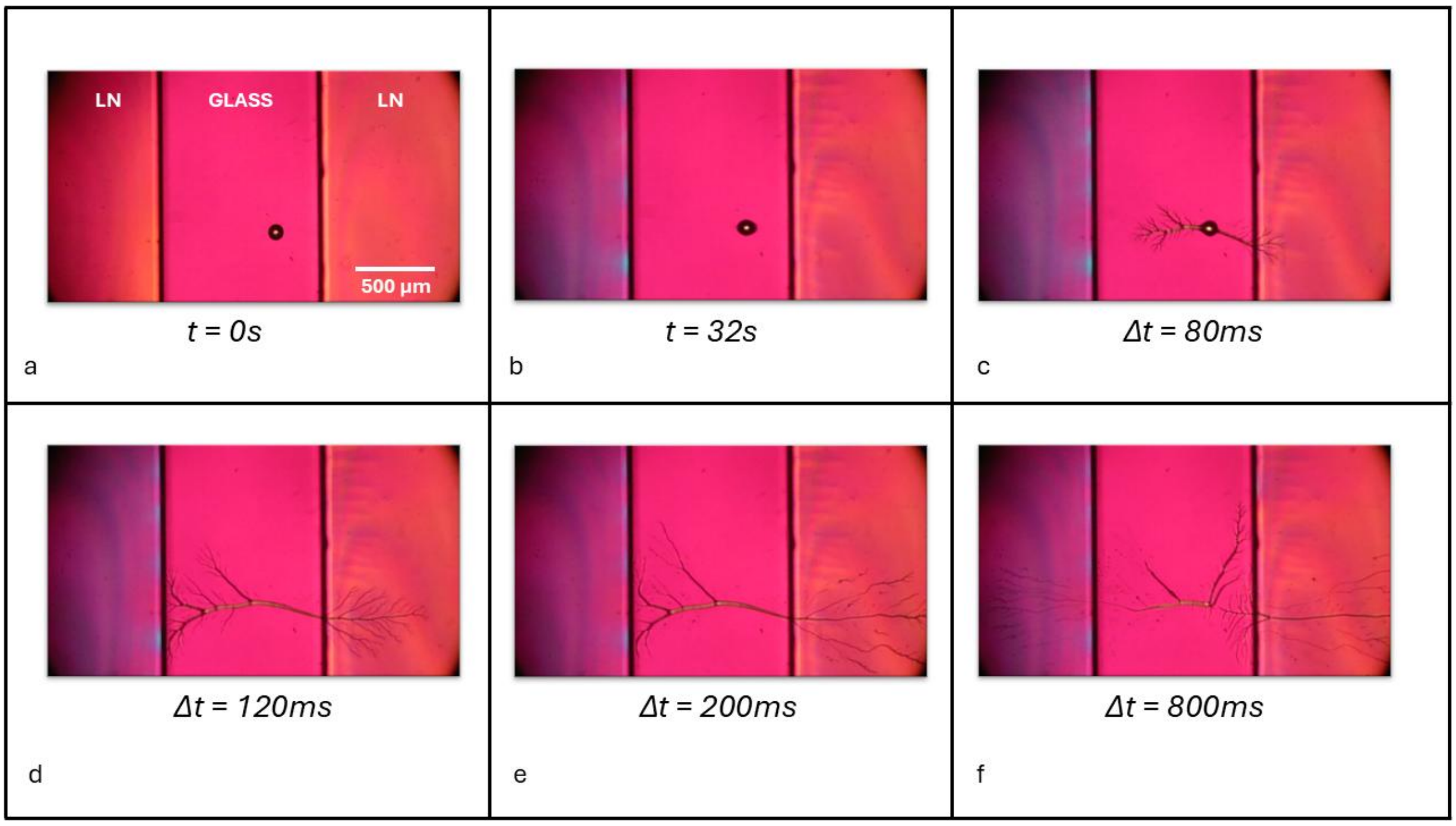

**Figure S9:** Frames describing the electromechanical instability of a $N_F$ sessile droplet positioned closer to one of the two LN crystals, when motion is not observed. a) Droplet's initial position. t=0 marks the beginning of the heating cycle; b) deformation and onset of electromechanical instability; c-e) the emitted fluid jets can travel in the region above the closest crystal and are repelled by the other; f) only when jets detach from the droplet, they become free to move over the previously forbidden region. The LN polar axes are both positive. $\Delta t$ in figures c-f indicates the time interval starting from fig. S7 b. Colors are due to the presence of a full wave plate (both the glass substrate and the z-cut LN crystals appear dark under crossed polarizers. The full wave plate increases the contrast of the images).

*Antiparallel polar axes*

When LN polar axes are antiparallel, droplet motion cannot be observed, regardless of droplet position with respect to the center of the gap, confirming that the electrostatic charging of $N_F$ droplets requires highly non-uniform electric fields.
Droplets exposed to this quasi-uniform electric field undergo significant, elastic deformation, that possibly, but not necessarily, ends up in droplets fission. The deformation is observable only in the $N_F$ phase and is largely reversible. An example is shown in Fig. S10 for a gap of 600 μm between the LN crystals. Figure S10 a shows the $N_F$ droplet right after the phase transition. Panels a, b and c are frames extracted during the cooling cycles and report the droplet's deformation. As the temperature variation stops, the droplet regains its original shape (d) and deforms again during the heating cycle (e,f).

It is interesting to note the presence of a tiny droplet on the right-hand side of panel a. This droplet is positioned outside the crystals gap above the LN with negative polar axis close to its border, in a region where the field is highly non-uniform. Comparing panel a to panel b, it appears clear that this droplet is moving toward the center of the LN crystal, where the uncompensated charge is negative upon cooling[5]. This is an additional confirmation that $N_F$ droplets can acquire an electrostatic charge only when they are exposed to a strongly non-uniform electric field and that this charge is positive.

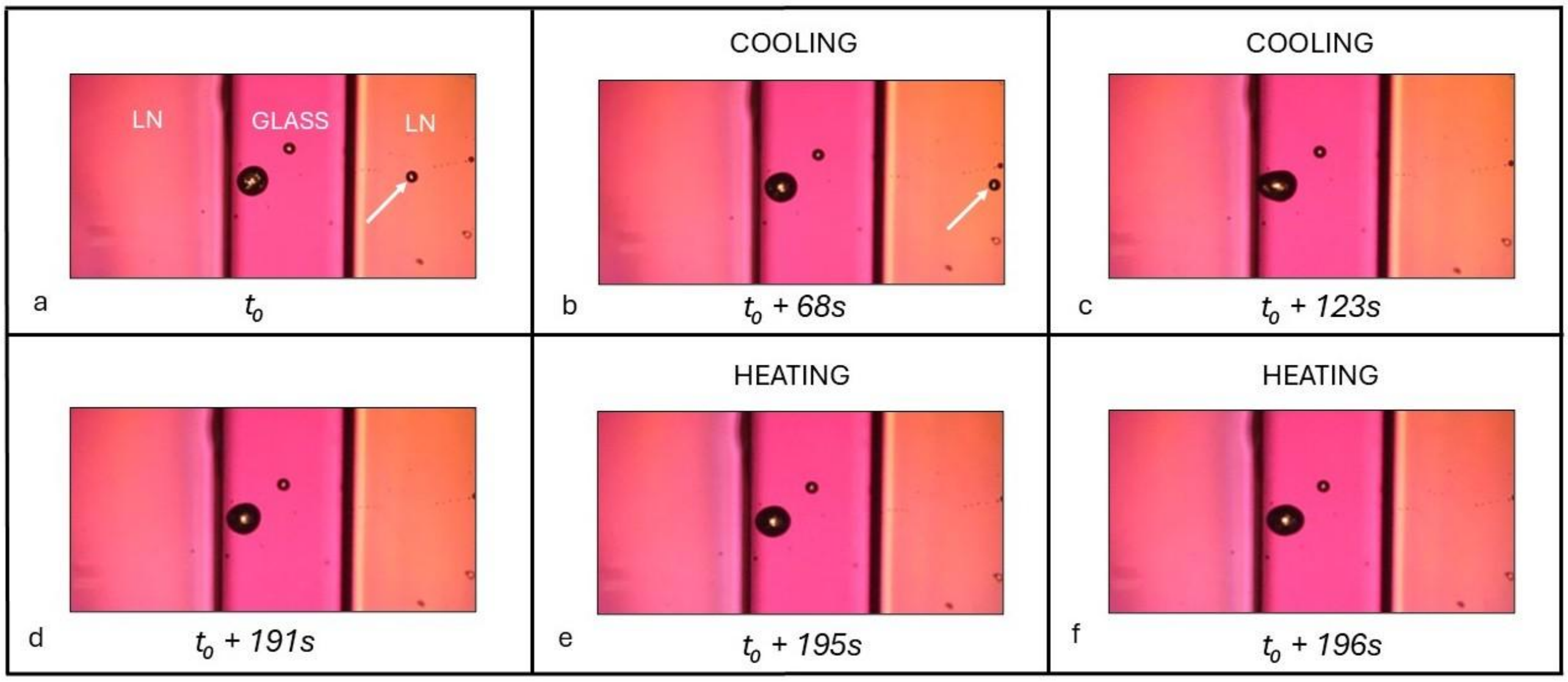

**Figure S10:** Droplet deformation during cooling and heating cycles in case of LN crystals with antiparallel polar axes (positive on the left and negative on the right). $t_0$ in a) marks the transition to the $N_F$ phase. b,c) droplet deformation during the cooling cycle; d) droplet shape after the end of the cooling cycle, at fixed temperature; e,f) further deformation during the heating cycle. Note the tiny droplet indicated by a white arrow, on the right-hand side of panel a. This droplet is positioned outside the crystals gap, in a region where the field is non-uniform and moves toward the center of the LN crystal with negative polar axis (b). In panel c) the droplet is no longer visible, being outside the field of view. Colors are due to the presence of a full wave plate.

The amount of droplet's deformation depends on the electric field squared. This is shown in Fig. S11, reporting the increase of droplets diameter $\Delta\phi$ as a function of the temperature variation $\Delta T$ for two different experiments, during the cooling cycle. $\Delta T$ is defined as the difference between the transition temperature and the actual temperature at a given time. Since the pyroelectric field is proportional to $\Delta T$[15], we can reasonably assume a linear relation between $\Delta T$ and the fringing field present in the gap between the crystals. Figure S11 thus indicates a quadratic dependence of droplets deformation on the square of the external electric field, resembling an electrostriction-like effect, i.e. a mechanical distortion due to electrical polarization[16]. The induced $\Delta\phi$ can exceed 20% of the initial diameter for values of the electric field of the order of $10^4 - 10^5$ V/m, the order of magnitude expected for the fringing field in our experimental conditions. Looking at this phenomenon as an electrostrictive-like effect and defining an effective electrostriction coefficient *a* as the ratio between the relative deformation ($\Delta\phi/\phi_0$) and electric field squared[17], one obtains the remarkable value $a \approx 10^{-10}$-$10^{-11}$ $m^2/V^2$.

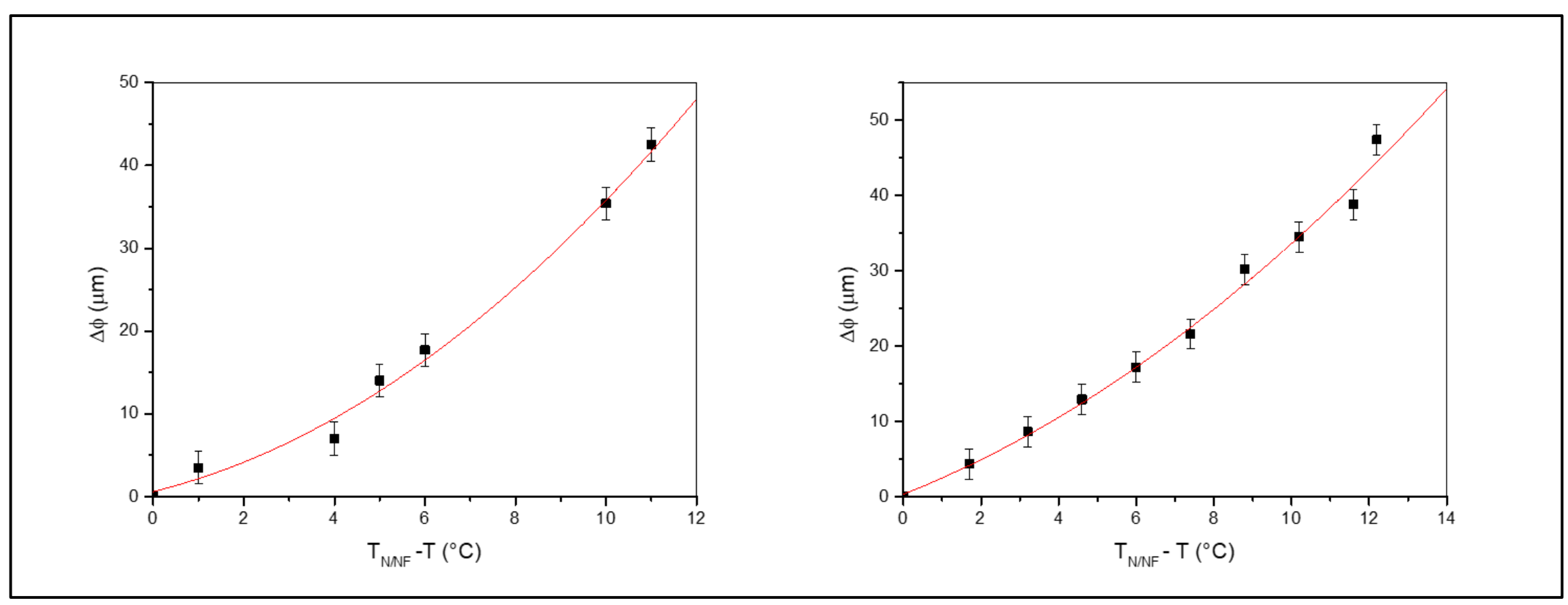

**Figure S11:** Increase of droplets diameter Δϕ as a function of the difference between the transition temperature to the $N_F$ phase and the actual temperature at a given time, for two different experiments, during the cooling cycle. The red lines are best fits of the experimental data, based on the function $y = Ax^2$.

**Videos**

**Video S1:** 140 μm RM734 droplet placed close to a LN crystal (on the right-hand side, marked as LN) with negative polar axis, subject to a cooling and a heating cycle. Cooling from 125°C (t = 0) to 105 °C (t= 16 s). The droplet moves toward the crystal until it exits the field of view. From t = 16 s to t = 23 s the temperature is kept fixed at 105°C. Upon heating the system from 105 °C (t =23 s) to 145°C (t = 29 s), the droplet moves in the opposite direction. Video has been sped up x10 in the first part (cooling) and x6 in the second part (heating). The initial distance between LN border and droplet center is 389 μm.

**Video S2:** 131 μm RM734 droplet placed close to a LN crystal (on the left-hand side, marked as LN) with positive polar axis, subject to a cooling and a heating cycle. Cooling from 125°C (t = 0) to 105 °C (t= 10 s). From t = 10 s to t = 15 s the temperature is kept fixed at 105°C. Upon heating the system from 105 °C (t =15 s) to 165°C (t = 26 s), the droplet moves in the opposite direction and stops a few seconds before the end of the thermal cycle. Video has been sped up x15 in the first part (cooling) and x6 in the second part (heating). The initial distance between LN border and droplet center is 1915 μm.

**Video S3-1:** 100 μm RM734 droplet placed close to a LN crystal (on the right-hand side, marked as LN) with positive polar axis, subject to a cooling and a heating cycle. Cooling from 125°C (t = 0) to 105 °C (t = 11 s). From t = 11 s to t = 12 s the temperature is kept fixed at 105°C. Upon heating the system from 105 °C (t = 12 s) to 140°C (t = 16 s), the droplet moves in the opposite direction until motion is stopped by an electromechanical instability event. Video has been accelerated x10 in the first part (cooling) and x6 in the second part (heating) up to the explosion, which is in real time.

**Video S3-2:** 262 μm RM734 droplet placed close to a LN crystal (on the right-hand side, marked as LN) with positive polar axis, subject to a cooling and a heating cycle. Cooling from 120°C (t = 0) to 105 °C (t = 12 s). The droplet moves away from the crystal until it exits the field of view. From t = 12 s to t = 15 s the temperature is kept fixed at 105°C. Upon heating the system from 105 °C (t = 15 s) to 140°C (t = 25 s), the droplet moves in the opposite direction, but the motion is stopped by an electromechanical instability event that involves first a smaller droplet at rest on the right. The motion of the secondary droplets can also be observed. Video has been sped up x10 in the first part (cooling) and x5 in the second part (heating), up to the explosion, which is in real time.

**Video S4:** 100 μm RM734 droplet placed close to a LN crystal (on the right-hand side, marked as LN) with positive polar axis, positioned above the imaginary green line in Fig. 1a' of the main text. Only the heating cycle is shown. Heating from 105 °C (t = 0) to 165 °C (t = 18 s). The droplet moves toward the LN crystal along a diagonal trajectory, following the field lines. Motion stops a few seconds before the end of the heating cycle, at the transition to the N phase ($T_{NF/N}$ = 150°C, t = 13 s). The motion of tiny secondary droplets is also visible. These droplets are the result of the explosion of a larger droplet originally positioned close to the 100 μm one. Video has been sped up x3.

**Video S5:** 178 μm RM734 droplet placed close to a LN crystal (on the right-hand side, marked as LN) with positive polar axis. Only the heating cycle is shown. Droplet motion toward the LN crystal is hindered by the fast variations of the fringing field, visualized as fluctuations of the LN birefringence, due to instabilities of the temperature controller and to piezoelectricity of the ferroelectric crystal. Heating from 105 °C (t=0) to 145 °C (t = 25 s).

**Video S6:** 10 μm thick RM734 cell with PTFE-coated glasses as confining surfaces. RM734 is in the $N_F$ phase. The cell is positioned close to a LN crystal (on the left-hand side of the screen). In the very first part of the video the system is cooled down. At t = 1 s, the sign of the temperature variation is reverted, which results in a field inversion and in a reorientation of **P** following the new direction of the field. Close to the end of the video, it is possible to observe the motion of the LC/air interface, most likely driven by electrocapillary flow of the charged $N_F$ interface under the LN fringing field.

**Video S7:** 133 μm RM734 droplet placed close to a LN crystal (on the left-hand side, marked as LN) with negative polar axis. Only the cooling cycle is shown, from 120°C (t = 0) to 105 °C (t = 18 s). During the first 8 seconds of motion, it is possible to observe the release of pieces of fluid in form of tiny droplets. Droplet speed increases after the release. The initial distance between LN border and droplet center is 360 μm. Video has been sped up with a decreasing acceleration, from x15 in the first part to x2 in the last part, so to capture properly the end of droplet motion at the end of the cooling cycle.